%% file: main.tex
\input{settings.tex} 
\input{tim_macros.tex}

\title{The Price of Anarchy in Markets for Lemons}
\author{Naveen Durvasula\thanks{Columbia University. Email:
\texttt{nkd2126@columbia.edu}. Supported in part by NSF award CNS-2212745.}
\and Tim Roughgarden\thanks{Institute for Advanced Study, Columbia
  University, and a16z. 
Email: \texttt{tim.roughgarden@gmail.com}. 
}}
\date{\today}

\begin{document}

\maketitle

\input{abstract.tex}

\newpage

\input{intro2.tex}

\input{model.tex}
\input{reduction.tex}
\input{retention.tex}

\input{twosided.tex}
\input{mechanisms.tex}

\appendix

\input{correlation_example.tex}
\input{appendix.tex}

\bibliographystyle{alpha}
\bibliography{refs}

\input{aimethodology.tex}

\end{document}

%% file: settings.tex
\documentclass[11pt]{article}
\usepackage[margin=1in]{geometry} 
\usepackage{amsmath}
\usepackage{tcolorbox}
\usepackage{amssymb}
\usepackage{amsthm}
\usepackage{mathtools}
\usepackage{lastpage}
\usepackage{fancyhdr}
\usepackage{accents}
\usepackage{systeme}
\usepackage{tikz}
\usepackage{enumerate}
\usepackage{tikz-cd}
\usepackage{enumitem}
\usepackage{bbm}

\newenvironment{amatrix}[1]{%
  \left[\begin{array}{@{}*{#1}{c}|c@{}}
}{%
  \end{array}\right]
}

\newcommand{\R}{\mathbb{R}}

\newcommand{\bra}[1]{\left[#1\right]}
\newcommand{\pa}[1]{\left(#1\right)}
\newcommand{\set}[1]{\left\{ #1 \right\}}
\newcommand{\abs}[1]{\left|#1\right|}

\newcommand{\pr}[1]{\Pr\bra{#1}}

\newcommand{\ex}[1]{\mathbb{E}\bra{#1}}
\newcommand{\exx}[2]{\mathbb{E}_{#1}\bra{#2}}

\newcommand{\supp}{\operatorname{supp}}
\newcommand{\mc}{\mathcal}

\newcommand{\argmax}{\operatorname{argmax}}
\newcommand{\argmin}{\operatorname{argmin}}

%% file: tim_macros.tex
\usepackage{epsfig}
\usepackage{amssymb}
\usepackage{fullpage}
\usepackage{amsmath}
\usepackage{ifthen}
\usepackage{verbatim}
\usepackage{hyperref}
\usepackage{xspace}
\usepackage{bm}
\usepackage{setspace}
\usepackage{tablefootnote}

\theoremstyle{plain}
\newtheorem{theorem}	 			{Theorem}[section]
\newtheorem{lemma}		[theorem]	{Lemma}

\theoremstyle{definition}

\newtheorem{definition}[theorem]{Definition}

\newtheorem{example}[theorem]{Example}

\newcommand{\prob}[2][]{\text{\bf Pr}\ifthenelse{\not\equal{}{#1}}{_{#1}}{}\!\left[#2\right]}
\newcommand{\expect}[2][]{\text{\bf E}\ifthenelse{\not\equal{}{#1}}{_{#1}}{}\!\left[#2\right]}

%% file: abstract.tex
\begin{abstract}
We study the worst-case equilibrium gains from trade in simple
bilateral trade mechanisms with interdependent valuations.
Equilibria in such settings are typically complex, 
as proposed prices can convey private information held by the proposer.
Such equilibria suffer lost
gains from trade both due to the inefficiency introduced by screening
(as in independent private value settings) and due to adverse
selection. We identify broad conditions under which, nevertheless,
sequential equilibria are guaranteed to achieve a constant fraction of
the first-best gains from trade. 
Our bounds quantify the central role of rare pessimistic seller
signals and rare optimistic buyer signals.
The results apply to
seller-proposer, buyer-proposer, and random-proposer mechanisms.
The key idea in our analysis is to relate equilibrium gains from trade to proposer profits in an auxiliary market with independent private values.
\end{abstract}


%% file: intro2.tex
\section{Introduction}

\subsection{Bilateral Trade}\label{ss:bl}

In the basic {\em bilateral trade problem}, there is one potential
seller, in possession of an item, and one potential buyer. 
The outcome of a bilateral trade mechanism is an assignment of the
item (which may remain with the seller or be trasnsferred to the
buyer) and a monetary transfer (typically, from the buyer to the
seller in exchange for the item). Ideally, trade would occur if and
only if the buyer's value for the item exceeds that of the
seller's. This ``first-best'' scenario would maximize the gains from trade.

\vspace{-.1in}

\paragraph{Seller-proposer mechanism.}
For example, one way to organize bilateral trade is through
the {\em seller-proposer mechanism}, which can
be viewed as a two-stage game:
\begin{itemize}

\item [] \textbf{Stage 1:} The seller chooses a proposed sale
  price~$p$ for their item.

\item [] \textbf{Stage 2: } The buyer chooses whether to accept the offer
  (paying~$p$ to
  receive the item) or not (paying nothing and receiving nothing).

\end{itemize}
How will the seller and buyer behave in this mechanism?
Informally, we expect each to maximize their expected utility, given
whatever information they have. Assume their utility functions are
{\em quasi-linear}, with utility equal to their value for the item in
their possession (if any) minus their payment (if any). The remaining
modeling questions are then: what are the buyer's and seller's values?;
and what information about these values is available to them when they
choose an action?

\vspace{-.1in}

\paragraph{Independent private values (IPV).}
In the commonly studied {\em independent private values (IPV)} version
of the problem, 
the seller's and buyer's values, $v_S$ and $v_B$, respectively, are
drawn independently from commonly known distributions $F_S$ and $F_B$.
Each agent initially knows their own value (with certainty) and
nothing about the other's value (other than the distribution from
which it is drawn). 

For example, in the first stage of the seller-proposer mechanism, the
seller chooses a price~$p$ knowing their own value~$v_S$ and the
distribution~$F_B$ from which the buyer's value is drawn. In the
second stage, the buyer chooses whether to accept the offer knowing
their own value~$v_B$, the distribution~$F_S$ from which the seller's
value is drawn, and the price~$p$ offered by the seller in the first
stage.  The optimal action for the buyer in the second stage, as a
function of the price~$p$ proposed by the seller in the first stage,
is clear. The buyer receives utility $v_B-p$ if they accept the offer
and~0 otherwise, and therefore will accept the offer if and only if
$v_B \ge p$.\footnote{For simplicity, we assume for now that a buyer
  will tie-break in favor of accepting.}
Given this characterization of optimal buyer behavior, the seller
then faces a {\em monopoly pricing} problem. The
optimal strategy here would also appear straightforward: choose a price~$p=p(v_S)$ that
maximizes the expected utility
\begin{equation}\label{eq:intro}
\underbrace{v_S \cdot \prob[v_B \sim F_B]{v_B < p}}_{\text{offer rejected}}
+ 
\underbrace{p \cdot \prob[v_B \sim F_B]{v_B \ge p}}_{\text{offer accepted}}
\end{equation}
One subtlety is that the maximizer of~\eqref{eq:intro} generally 
reveals partial or even full information about the seller's
value~$v_S$ to the buyer.
But {\em because we are currently in the IPV settting},
the buyer's value (and hence its optimal second-stage action) is
independent of the seller's value. Thus, the buyer's second-stage
choice above really is a best response to the seller choosing the
maximizer of~\eqref{eq:intro} in the first stage
(and vice versa).\footnote{This would remain true if the seller's and
  buyer's values were correlated, provided they were still
  private.}\textsuperscript{,}\footnote{The precise equilibrium
  concept we consider corresponds to Kreps and
  Wilson~\cite{krepswilson1982} sequential equilibria
and is discussed at length
  in Section~\ref{ss:eq}. Such a refinement is necessary, as
  otherwise there are always trivial ``no-trade'' (Bayesian-Nash) equilibria.\label{foot:bne}}

\vspace{-.1in}

\paragraph{Equilibrium gains from trade.}
In the equilibrium above, there are cases in which a beneficial trade
is not realized (when $v_B \in (v_S,p)$ for $p$
maximizing~\eqref{eq:intro}). That is, some of the potential gains
from trade are lost due to screening by the monopolist seller.
This is expected: the Myerson-Satterthwaite theorem~\cite{MS83}
states that, under weak assumptions (which are satisfied in the
present setup), even in the IPV setting, no weakly budget-balanced
mechanism (such as the seller-proposer mechanism)
can achieve the first-best gains from trade.

\subsection{Interdependent Values and Asymmetric Information}


\paragraph{Correlated values.}
For many applications, the IPV setting is an inappropriate assumption.
For example, the item in question may have an intrinsic ``quality''
that correlates the seller's and buyer's values (e.g., with each an
increasing function of the item's quality). Akerlof~\cite{akerlof1970}
proposes the used car market as one paradigmatic example (with a car
in good condition worth much more to both parties than a lemon);
Milgrom and Weber~\cite{milgrom_weber} use mineral rights as their
leading example (e.g., with a plot of land worth a lot to all of
the participating oil companies if it harbors oil, and worth little to
all if not).

\vspace{-.1in}

\paragraph{Partial information.}
A second naturally arising complication is that some or all parties may
have only {\em partial} information about their values for the item,
for example if the item's quality is uncertain. In Akerlof's canonical
``market for lemons''~\cite{akerlof1970}, the seller is strictly more
informed than the buyer: the seller knows their car's quality with
certainty, while the buyer does not (knowing only a prior
distribution over car qualities). In the Milgrom-Weber mineral rights
example~\cite{milgrom_weber}, all of the participants generally have
only partial information about item quality (i.e., the presence of
oil), perhaps derived from the assessments of their in-house
geologists.

\vspace{-.1in}

\paragraph{Interdependent values.}
Milgrom and Weber~\cite{milgrom_weber} introduced the {\em
  interdependent values model} as the canonical framework for
reasoning about applications in which
there is a common value component (e.g., item quality) and/or
uncertainty about one's value (e.g., due to uncertainty about item
quality). We state the model precisely in Section~\ref{sec:model}. Here,
we observe that even in the seemingly near-trivial setup 
of the seller-proposer mechanism for bilteral trade, very simple
special cases of interdependent values already lead to complex
equilibrium behavior. The following example already illustrates many
of the subtleties that arise in our general model. The example also
showcases a second source of inefficiency that arises with
interdependent values (beyond the lost gains from trade from screening
seen already in the IPV setting), namely {\em adverse selection}.

\begin{example}[Pooling at Equilibrium with Interdependent
  Values]\label{ex:pooling}
\mbox{}\\

\vspace{-.5\baselineskip}
\noindent
\textbf{The setup:} Consider the seller-proposer mechanism for bilateral trade from
Section~\ref{ss:bl}. Suppose now that the seller's item has a {\em
  quality}~$q$, which is equally likely to be~1 or~2. The extents to
which the seller
and buyer care about quality, $a_S$ and $a_B$, respectively, are
private and drawn from the uniform distribution on~$[0,1]$
(independently of each other and of~$q$).
The
induced seller and buyer values are then $v_S := a_S \cdot q$ and $v_B
:= a_B \cdot q$, respectively. 
The same value (e.g., $v_S =
\tfrac{1}{2}$) may arise for multiple reasons, such as a high quality
item ($q=2$) coupled with a relatively low private type ($a_S =
\tfrac{1}{4}$) or a low quality item ($q=1$) coupled with an average
private type ($a_S = \tfrac{1}{2}$).

\vspace{.5\baselineskip}
\noindent
\textbf{The first-best benchmark:}
There are positive gains from trade if and only if $a_B > a_S$, and
the expected maximum gains from trade are $\expect{q} \cdot
\expect{(a_B-a_S)_+} = \tfrac{3}{2} \cdot \tfrac{1}{6} = \tfrac{1}{4}$.

\vspace{.5\baselineskip}
\noindent
\textbf{The information structure:}
Suppose that each agent is fully informed about
their own private type ($a_B$ or $a_S$) but uninformed about the
other's (knowing only its distribution).
In the spirit of Akerlof~\cite{akerlof1970}, there is
asymmetric 
information: the seller is fully informed (i.e., knows $q$ with
certainty) while the buyer is uninformed (i.e., knows only $q$'s prior
distribution).

\vspace{.5\baselineskip}
\noindent
\textbf{Optimal buyer behavior:}
In the second stage of the seller-proposer mechanism, 
given the seller's strategy in the first stage, the
optimal strategy for the buyer in the second stage is: compute the
expected quality $\expect{q}$ of the item (given the prior
distributions, the seller's strategy, and the offered price $p$), and
accept the seller's offer if and only $a_B \cdot \expect{q} \ge p$.
Because the price~$p$ may reveal
information about the seller's value, and thus about $q$,
%
%
%
the best response of the buyer depends
on the seller's strategy (and not just the offered price). 

\vspace{.5\baselineskip}
\noindent
\textbf{Seller behavior at equilibrium (failed attempt):}
A seller with a high-quality item would like to signal that quality to
the buyer, so that the buyer will value it highly and be willing to
pay a relatively high price. Unfortunately for the seller, they can
only communicate item quality indirectly, through the price that they
offer.

To derive a possible equilibrium seller strategy,
let's tentatively assume that the seller chooses its
price~$p=p(a_S,q)$ as a
function only of its value $v_S = a_S \cdot q$. Let's further assume
that the seller's price will reveal to the buyer whether $v_S \in [0,1]$
or $v_S \in (1,2]$. It is then easy to reason about the downstream
behavior of the buyer. 
If the price~$p$ reveals that~$v_S > 1$, the buyer learns with
certainty that~$q=2$, and accepts the offer if and only if
$2a_B \ge p$.
If the price~$p$ reveals that~$v_S \le 1$, the buyer's posterior
beliefs assign a probability of~$\tfrac{2}{3}$ to the event that
$q=1$, and the buyer therefore accepts the offer if and only if
$\tfrac{4}{3} \cdot a_B \ge p$. Choosing monopoly
prices (i.e., prices that maximize~\eqref{eq:intro}) separately on
each interval yields the following seller strategy:
\begin{equation}\label{eq:strat0}
	p(v_S) := \begin{cases}
		\frac23 + \frac12 v_S & \text{if $v_S \in [0,1]$}\\
		1 + \frac12 v_S & \text{if $v_S \in (1,2]$.}
	\end{cases} 
\end{equation}
The best response of the buyer to the seller strategy
in~\eqref{eq:strat0} is react to prices in
$[\tfrac{2}{3},\tfrac{7}{6}]$ and $(\tfrac{3}{2},2]$ with a valuation
of $\tfrac{4}{3}a_B$ and $2a_B$, respectively. But holding this buyer
strategy fixed, is this seller strategy a best response in return? \textit{No:}
a seller with value~$v_S=1$ could earn higher expected utility
($\approx \tfrac{9}{8}$ vs.\ $\tfrac{49}{48}$) by offering a price of
$\approx \tfrac{3}{2}$ rather than~$\tfrac{7}{6}$, in effect tricking
the buyer into believing that~$q=2$ when that may not have been the
case. Put differently, a seller cannot credibly commit to choosing
prices according to~\eqref{eq:strat0}, because high-value sellers of
low-quality items would be incentivized to impersonate medium-value sellers
of high-quality items. 

\vspace{.5\baselineskip}
\noindent
\textbf{Seller behavior at equilibrium (corrected):}
At equilibrium, a price intended to reveal that~$q=2$ must be
sufficiently large to deter such impersonations, so that buyers can be
sure they are protected from adverse selection. This leads to
the following revised seller strategy:
\begin{equation}\label{eq:strat1}
	p(v_S) := \begin{cases}
		\frac23 + \frac12 v_S & \text{if $v_S \in [0,1]$}\\
		\frac{3 + \sqrt{5/6}}{2} & \text{if $v_S \in (1, 1 + \sqrt{5/6}]$}\\
		1 + \frac12 v_S & \text{if $v_S \in [1+\sqrt{5/6},2]$.}
	\end{cases} 
\end{equation}
With buyer behavior as before, the expected utility of the marginal
low-value seller (i.e., $v_S=1$) is the same ($\tfrac{49}{48}$)
whether it offers a price of~$\tfrac{7}{6}$ or $\frac{3 +
  \sqrt{5/6}}{2}$, so impersonation is no longer profitable (and is
strictly unprofitable for lower-value sellers).
These seller and buyer strategies therefore constitute an
equilibrium.\footnote{To fully specify the equilibrium, one must
  define the buyer's behavior for all possible prices, including those
  in the interval $(\tfrac{7}{6},\frac{3 + \sqrt{5/6}}{2})$. This
  requires specifying ``off-path beliefs'' for the buyer; see
  Section~\ref{sec:model} for details.}
In~\eqref{eq:strat1}, sellers with value in the interval $(1, 1 +
\sqrt{5/6}])$ form a ``pool,'' all offering the same price. These sellers
would prefer to offer lower prices to increase the likelihood of trade
(holding the buyer strategy fixed),
but they cannot credibly do so; gains from trade suffer as a result.
A calculation 
shows that, in this
equilibrium, the expected gains from trade are~$\tfrac{1}{6}$, or
two-thirds of the first-best benchmark.\footnote{There are more complex
equilibria that achieve only~$\tfrac{1}{8}$ of the expected maximum
gains from trade, and an ad hoc argument shows that these are the
worst equilibria in this example.}
\end{example}

\subsection{A Barrier to Large Gains from Trade: Rare Pessimistic Signals}

In Example~\ref{ex:pooling}, 
equilibria of the seller-proposer
mechanism fail to achieve the full first-best gains from trade, but
they do all approximate that benchmark up to a constant factor.
Could such a guarantee hold
in general for the seller-proposer mechanism?

\begin{example}[Rare Signals Can Limit Equilibrium Gains from Trade]\label{ex:rare}
\mbox{}\\

\vspace{-.5\baselineskip}
\noindent
\textbf{The setup:} 
We make only small changes to Example~\ref{ex:pooling}.
Private types $a_S,a_B$ are now drawn independently and uniformly from
$[\tfrac{1}{2},1]$ rather than~$[0,1]$.
The item quality~$q$ is now drawn uniformly from~$[0,1]$ rather than
from~$\{1,2\}$ (again, independently of~$a_S$ and $a_B$).

\vspace{.5\baselineskip}
\noindent
\textbf{The first-best benchmark:}
The expected maximum gains from trade are now $\expect{q} \cdot
\expect{(a_B-a_S)_+} = \tfrac{1}{2} \cdot \tfrac{1}{12} = \tfrac{1}{24}$.

\vspace{.5\baselineskip}
\noindent
\textbf{The information structure:}
The example is now parameterized by a threshold
$t \in (0,\tfrac{1}{2}]$.
The seller no longer knows~$q$ with certainty, only a trustworthy
``signal'' $\sigma$ that indicates whether $q \le t$ (the ``low''
signal $\sigma = L$) or $q > t$ (the ``high'' signal $\sigma = H$).
The buyer is uniformed (knowing only~$q$'s distirbution), as before.

\vspace{.5\baselineskip}
\noindent
\textbf{An equilibrium:}
In one equilibrium in this example, the seller's strategy fully
reveals its signal, with a high-signal seller
refusing to sell (i.e., setting~$p=+\infty$) and a low-signal seller
offering a suitable monopoly price (as in~\eqref{eq:intro}).
In the latter case, because the signal is revealed to the buyer and
$\expect{q \,|\, \sigma = L } = \tfrac{t}{2}$, the buyer will accept
the offered price~$p$ if and only if $\frac{t}{2} \cdot a_B \ge
p$. The low-signal seller will set the price~$p$ to
maximize~\eqref{eq:intro}, with $v_S$ identified with the seller's
expected value $\tfrac{t}{2} \cdot a_S$ given its signal $\sigma = L$,
and with the buyer value distribution~$F_B$ identified with the
uniform distribution on $[\tfrac{t}{4},\tfrac{t}{2}]$. That is:
\begin{equation}\label{eq:rare}
	p(a_S,\sigma) := \begin{cases}
		+\infty & \text{if $\sigma = H$}\\
		\tfrac{t}{4} \cdot (1 + a_S) & \text{if $\sigma = L$.}
	\end{cases} 
\end{equation}
Because a low-signal seller has value $\tfrac{t}{2} \cdot a_S \le
\tfrac{t}{4} \cdot (1+a_S)$, they have nonnegative gains from trade
under the strategy in~\eqref{eq:rare} and cannot benefit from
impersonating a high-signal seller. 
A high-signal seller has value $\expect{q \,|\, \sigma = H }
  \cdot a_S = \tfrac{1+t}{2} \cdot a_S \ge \tfrac{t}{2}$
for the item and cannot benefit from impersonating any low-value
seller.\footnote{Again, fully specifying the equilibrium requires
  specifying beliefs and actions for the buyer for prices outside of
  $[\tfrac{3}{8}t,\tfrac{t}{2}] \cup \{+\infty\}$; see
  Section~\ref{ss:eq} for details.}
This equilibrium mirrors the unraveling Akerlof~\cite{akerlof1970}
identifies in his markets for lemons: buyers assume adverse selection
and act as if the only
willing sellers are those receiving a low signal, and this assumption
is self-reinforcing (with high-signal sellers unable to charge any
price at which they would be willing to trade).

\vspace{.5\baselineskip}
\noindent
\textbf{Equilibrium gains from trade:}
Because trade occurs only when the seller receives a low signal (which
occurs with probability~$t$), and the expected quality~$q$ is
only~$t/2$ in this case, the expected gains from trade at this
equilibrium are~$O(t^2)$. Thus, as the signal indicating
low quality gets increasingly rare, the expected gains from trade
realized at this equilibrium are a vanishingly small fraction of the
first-best benchmark.\footnote{There are also other equilibria in this
  example for which the expected gains from trade are a constant
  fraction of the first-best benchmark (independent of~$t$). Our focus
  in this paper is on the gains from trade in worst-case equilibria.}

\end{example}

\input{overview.tex}

\input{related_work.tex}

%% file: overview.tex
\subsection{Proof Approach}\label{ss:approach}

Example~\ref{ex:rare} shows that rare pessimistic signals are an
obstruction to good worst-case approximation bounds on equilibrium
gains from trade in the seller-proposer mechanism. Could it be that
there are no other obstructions? Our results, summarized in
Section~\ref{ss:results}, provide affirmative answers to this and
related questions. In the service of fully describing and interpreting
those results, this section outlines a key reduction step in our
modular proof approach. 

Suppose we retain the same setting we have been considering thus
far. An informed seller proposes to an uninformed buyer. The seller
and buyer have values $v_S := a_S \cdot q$ and $v_B := a_B \cdot q$
for the item respectively, where $a_S$ an $a_B$ are private and
independent draws from a distribution~$\mc G$ with nonnegative
support, and $q$ is drawn independently from a distribution $\mc
F$.
The seller's information is given by their private observation of a
signal $\Sigma_S$ pertaining to the realization of $q$.

As we've seen in Examples \ref{ex:pooling} and \ref{ex:rare},
equilibrium strategies can be very complex even in simple settings.
Finding an equilibrium may require reasoning about pooling strategies,
imitation, and more broadly, information about the item's quality that
is communicated by the proposed price and the consequences for the
buyer's equilibrium behavior.
In the spirit of smoothness arguments~\cite{roughgarden2015intrinsic},
our approach sidesteps the challenge of directly characterizing
equilibria by considering deviations that could be made by a player in
any equilibrium. By best-response optimality at equilibrium, the
player's equilibrium payoff must at least be their deviation payoff.
We specifically ask:
{\em What profit can the seller guarantee without knowing how their
  proposed price will impact the buyer's estimation of item quality?}

Recall that the proposed price may be used by the buyer to infer
information known to the seller that is relevant to the buyer's
valuation, resulting in some posterior belief by the buyer over the
signal that the seller observed.  But no matter how the buyer
interprets the proposed price, their valuation must be at least be as
high as what it would be if the buyer somehow knew that the seller
observed their most pessimistic signal. Concretely, define
$\sigma_S^-$ to be a maximally pessimistic signal and $d$ to be the
expected quality of the item given this signal realization:
\begin{equation}
	\sigma_S^- \in \argmin_{\sigma_S \in \supp \Sigma_S} \ex{q \mid \Sigma_S = \sigma_S}, \qquad d:= \ex{q \mid \Sigma_S = \sigma_S^-}.
\end{equation}
A buyer whose type is $a_B$ must have a valuation $v_B \ge d a_B$
regardless of how they revise their estimate for item quality given
the seller's proposed price. Given sequential rationality of the
buyer, it follows that any seller who proposes a price $p$ has a
guarantee that with probability at least $\pr{d a_B \ge p}$, the buyer
will accept.\footnote{As noted in footnote~\ref{foot:bne}, sequential rationality is
  crucial for the existence of non-trivial worst-case equilibrium
  guarantees.}
In particular, if we consider a seller with type $a_S$ who observes $\Sigma_S = \sigma_S^-$ (and therefore estimates the item's quality to be $d$), proposing a price of $p > d a_S$ must guarantee at least 
\begin{equation}\label{eq:ipv}
	d a_S \cdot  \prob[a_B \sim \mc G]{d a_B < p}
+ 
p \cdot \prob[a_B \sim \mc G]{d a_B \ge p}
\end{equation} 
in expected utility. Notice the similarity of this payoff to
\eqref{eq:intro}: our key observation is that it is exactly the same
expected payoff a seller would receive in an {\em IPV market} in which
both the seller and buyer publicly observed the item's quality as
$d$. That is, we can lower-bound equilibrium gains from trade in the
original market by the proposing seller's profit in an IPV market
in which seller and buyer valuations are given by $\hat v_S := d a_S$ and
$\hat v_B := d a_B$, respectively. First-best gains from trade in this
IPV market are then given by
\begin{equation}\label{eq:R}
	R := \ex{(\hat v_B - \hat v_S)_+} = d\cdot \exx{a_B,a_S\underset{i.i.d}{\sim}\mc G}{\pa{a_B - a_S}_+}.
\end{equation}  
Our high-level proof approach is organized around this reduction. It consists of three main steps:

\vspace{-.1in}
\paragraph{Deviation profit in the IPV market.} We first show that
equilibrium gains from trade in the original market are at least the
seller's profit in the IPV market multiplied by the probability
$\prob{\Sigma_S = \sigma_S^-}$ that the most pessimistic signal is
realized.  The factor of~$\prob{\Sigma_S = \sigma_S^-}$ loss arises
because a seller that observes a signal other than the most
pessimistic one will assess the item's expected quality as more
than~$d$; the maximizer of
the revised expression~\eqref{eq:ipv}, with $da_S$ replaced by a
larger number, might result in no possibility of trade (cf., Example~\ref{ex:rare}).

\vspace{-.1in}
\paragraph{Screening Loss.} Next, under appropriate regularity
conditions on buyer and seller valuations, we show that the seller's
profit in the induced IPV market is a constant fraction of the first-best
gains from trade in that market. 
Passing from an interdependent setting to an IPV setting 
allows us to take advantage of the existing literature on
screening inequalities for such settings.

\vspace{-.1in}
\paragraph{IPV Benchmark Loss.} Finally, under appropriate regularity
conditions on the item quality distribution,
we show that the first-best gains from trade in the induced IPV market
can be related to those in the original market, losing another
(necessary) factor of~$\prob{\Sigma_S = \sigma_S^-}$.

\vspace{.75\baselineskip}

We generalize our results to markets in which both the buyer and
seller can be partially or fully informed about item quality.  
With an informed buyer, the seller must now also take into account
adverse selection (with the acceptance of an offer causing the seller
to revise its beliefs about quality upward).
Thus, the deviation payoff in \eqref{eq:ipv} is no longer valid,
because the seller's value for keeping the asset (given by the first
term in the sum) may change if the buyer chooses to accept. See
Section \ref{sec:two-sided} for further exposition on this
phenomenon.

Despite this challenge, we can adapt the same high-level proof
strategy by using a different definition for $d$. This comes from the
observation that \eqref{eq:ipv} is only an invalid lower-bound on
expected seller payoff due to potential upwards revisions in the
seller's estimate for the quality of the item conditioned on buyer
acceptance of an offer. We can therefore maintain a similar lower
bound by setting $d$ to be the expected quality of the item
conditioned on simultaneous observation of (i) the most pessimistic
signal by the seller, and (ii) the most optimistic signal by the
buyer. By conditioning on (i), we lower-bound the probability of buyer
acceptance, as before. By conditioning on (ii), we upper-bound the
seller's value for keeping the asset. By conditioning on both of these
events simultaneously, we recover the same style of guaranteed
deviation profit in terms of seller profit in an IPV market where both
the buyer and seller publicly observe the item's quality as $d$.

\subsection{Summary of Results}\label{ss:results}

\subsubsection{A Representative Special Case}

We now return to the original question we asked at the beginning of
Section \ref{ss:approach}: are rare pessimistic signals the only
obstruction to good worst-case approximation bounds on equilibrium
gains from trade when an informed seller proposes to an uninformed
buyer?  For a market instance $\mc I := (\mc F, \Sigma_S, \mc G)$,
where $\mc F$ denotes the distribution over item quality $q$,
$\Sigma_S$ denotes the informed seller's signal (which is generally
correlated with~$q$), and $\mc G$ denotes the marginal distribution
over the independent and private idiosyncratic components $a_S, a_B$
of the seller and buyer's valuations respectively, let
\begin{equation}
	W_S^-(\mc I) := \inf_{\text{Sequential Equilibria}}\frac{\text{Equilibrium Gains from Trade}}{\mathrm{FB}},
\end{equation}
where 
\begin{equation}
	\mathrm{FB} := \ex{(q a_B - q a_S)_+}
\end{equation}
denotes first-best gains from trade in the market. The $S$ in the subscript of $W_S^-$ denotes the seller-proposing mechanism. We make our precise equilibrium concept clear in Section \ref{sec:model}. At a high level, we consider a continuous-type adaptation of sequential equilibrium in the style of Kreps and Wilson \cite{krepswilson1982}. Crucially, this equilibrium concept imposes a well-defined notion of sequential rationality for the buyer.

By carrying out the proof approach outlined in Section
\ref{ss:approach}, we first show that if $\mc G$ has log-concave
density, then $W_S^-(\mc I)$ is controlled by the rarity of the most
pessimistic signal and the fraction $\frac{R}{\mathrm{FB}}$, where $R$
is the first-best gains from trade in our IPV reduction, as defined in
\eqref{eq:R}:\footnote{%
The set of all $\mc G$ with log-concave density encompasses a broad
family of distributions, including uniform, Gaussian, exponential, and
Laplace distributions. Shifting or truncating the distribution
maintains log-concavity of the density, and arbitrary
(nonzero) linear combinations of independent log-concave variables are
also log-concave \cite[Propositions 3.1 and 3.5]{saumardwellner2014}. 
}\textsuperscript{,}\footnote{Without any regularity conditions on
  valuations, the loss to
gains from trade already from screening and already in an IPV market
can be arbitrarily
high. Consider, e.g., the equal revenue distribution example
  given in footnote 2 of \cite{B+21}. This example can further be
  adapted so that seller and buyer valuations are identically
  distributed.}.\\

\noindent \textbf{Informal Theorem 1:} \textit{If the density of $\mc G$ is log-concave, then \[W_S^-(\mc I) \ge \frac{\prob{\Sigma_S = \sigma_S^-}}{e} \cdot \frac{R}{\mathrm{FB}}\]}.

This informal theorem is implied by our Theorem \ref{thm:reduction} in Section \ref{sec:reduction}.

Assuming appropriate regularity of the distribution over the item
quality $q$, the
probability of the most pessimistic signal 
also controls the fraction $\frac{R}{\mathrm{FB}}$ (and therefore
$W_S^-$).
The regularity condition is that $q \ge 0$ is dominated in convex order by an exponential variable of the same mean.\footnote{That is, if we let $Z$ be an exponential variable with mean $\exx{q \sim \mc F}{q}$, then for every convex function $\phi: \R \to \R$, 
$
	\exx{q \sim \mc F}{\phi(q)} \le \ex{\phi(Z)}.
$
This class of distributions includes, for example: 
%
Every MHR distribution (and therefore every distribution with log-concave density) \cite[Theorem~3.A.55]{shaked2007stochastic};
Every distribution $\mc F$ supported on the interval $[(1 - \log 2) \exx{q \sim \mc F}{q}, (1 + \log 2) \exx{q \sim \mc F}{q}]$; \cite[Theorem~3.A.4]{shaked2007stochastic}
Every mixture of distributions that each satisfy the condition and have the same mean \cite[Theorem~3.A.12(b)]{shaked2007stochastic};
Every (deterministic) convex combination of variables that satisfy the
condition, even if those variables are dependent.
(The last fact follows straightforwardly from Jensen's inequality.)
}

To state our informal result, we put these assumptions together to define a \textit{$\pi^-$-admissible instance};

\begin{definition}[$\pi^-$-admissible instance]
	A market instance $\mc I = (\mc F, \Sigma_S, \mc G)$ is {\em $\pi^-$-admissible} if:
	\begin{enumerate}
		\item $\mc G$ has log-concave density;
		\item $\mc F$ is dominated in convex order by an
                  exponential variable of the same mean; and
		\item The rarity of the most pessimistic signal is at least $\pi^-$: $\pr{\Sigma_S = \sigma_S^-} \ge \pi^-$.
	\end{enumerate}
\end{definition}

We then prove:\\

\noindent \textbf{Informal Theorem 2:} \textit{For every $\pi^-$-admissible instance $\mc I$, \[W_S^-(\mc I) \ge \frac{(\pi^-)^2}{4e}.\]}

We therefore answer our original question in the affirmative, and show
further a quadratic dependence on the likelihood of the most pessimistic
signal $\sigma_S^-$ in the worst case. Recall that
Example~\ref{ex:rare} shows that this quadratic dependence on $\pi^-$ is tight. This result is implied by our Theorem \ref{thm:retention}.

\subsubsection{Generalizations}

We generalize this result across three different axes: valuation structures, information structures, and mechanisms. We correspondingly augment our definition of a market instance $\mc I$ to be a tuple $(v, \mc F, \Sigma_S, \Sigma_B, \mc G)$. 

\paragraph{Affine Valuations.} The first term in the instance
definition, $v$, determines how the idiosyncratic component and common
component combine in the seller and buyer valuations. We now go beyond
multiplicative valuations to affine valuations
\begin{equation}
	v(c, x, y) := cx + y.
\end{equation}
Here, $c$ is the \textit{common component} to seller and buyer
valuations (we formerly called this $q$ for item quality) and $(x, y)$
form the idiosyncratic component. 
For example, we may model
bilateral trade in capital markets by considering $c$ to be the value
of a future cash flow of the item and $x$ to capture the effect of the
seller's or buyer's discount rate.
The additive term $y$ can further model some types of risk
aversion or private holding costs. Further discussion is given
in Section \ref{sec:model}.

We allow $x$ and $y$ to be dependent. So that screening losses remain
bounded, we extend our log-concave density assumption from~$x$ to the
joint distribution of $(x,y)$. 
With affine valuations, a new technical issue arises: perfect negative
correlation between~$x$ and~$y$ can cause preferences to collapse to a
singular distribution in expectation, which can ultimately cause
vanishing gains from trade at equilibrium relative to first-best. We
show this explicitly in Example~\ref{ex:correlation} in
Appendix~\ref{app:correlation}.  This example further shows the
tightness of our dependence on correlation in our bound below.

\paragraph{Two-Sided Information.} 
We also consider general market instances in which both the seller and
buyer may be differently informed. We allow the seller to observe a
signal $\Sigma_S$ and the buyer to observe a signal $\Sigma_B$. 
We assume that there are only a finite number of possible signal
realizations, and that signals are consistently ordered (e.g., with the
seller having a well-defined ``most pessimistic'' signal, independent
of the buyer's signal).
As telegraphed
in Section \ref{ss:approach}, 
with two-sided information, 
the likelihood of the buyer's most optimistic signal $\sigma_B^+$ also
plays a role.%
\footnote{Our result here directly generalizes the
  previous one: when the buyer is uninformed, the buyer observes
  their most optimistic (null) signal with probability one.}

To state our two-sided information result, we first define $(\pi^-, \pi^+, \rho)$-admissible market instances:
\begin{definition}[$(\pi^-, \pi^+, \rho)$-admissible instance]
	A market instance $\mc I = (\mc F, \Sigma_S, \mc G)$ is
{\em $(\pi^-, \pi^+,\rho)$-admissible} if 
	\begin{enumerate}
		
        \item $(x, y)$ has joint log-concave density under $\mc G$. If
          one coordinate is a constant, for example, with purely
          multiplicative values ($y = 0$) or purely additive values
          ($x = 1$), then the non-constant coordinate must have a
          log-concave density.

	\item The correlation of $x$ and $y$ is at least~$\rho$:
$\operatorname{Corr}(x, y) \ge \rho \in [-1,1]$.
		\item $\mc F$ is dominated in convex order by an exponential variable of the same mean.
		\item The rarity of the most pessimistic seller signal
                  is at least $\pi^-$ given any buyer signal
                  realization: $\min_{\sigma_B \in \supp
                    \Sigma_B}\pr{\Sigma_S = \sigma_S^- \mid \Sigma_B =
                    \sigma_B} \ge \pi^-$.
		\item The rarity of the most optimistic buyer signal
                  is at least $\pi^+$ given any seller signal
                  realization: $\min_{\sigma_S \in \supp
                    \Sigma_S}\pr{\Sigma_B = \sigma_B^+ \mid \Sigma_S = \sigma_S} \ge \pi^+$.
	\end{enumerate}
\end{definition}

We then prove:\\

\noindent \textbf{Informal Theorem 3:} \textit{For any $(\pi^-, \pi^+, \rho)$-admissible instance $\mc I$, \[W_S^-(\mc I) \ge \frac{(\pi^-)^2\pi^+ \sqrt{1 + \rho}}{4e}.\]}

While there is a quadratic dependence on $\pi^-$, there is only a linear dependence on $\pi^+$. This is because pessimistic seller signals cause trade to further take place in less valuable realizations of the common component. Optimistic buyer signals cause no such issue. This result is implied by our Theorem \ref{thm:two-retention}.

\paragraph{Alternate Mechanisms.} In addition to the seller-proposing
mechanism, we consider both the buyer-proposing mechanism and the
random-offerer mechanism. For a fixed market instance, the worst-case
equilibrium gains from trade will generally be different across these
three different mechanisms. Nonetheless, 
we obtain an identical worst-case (over instances) bound as in Informal
Theorem 3 for the buyer-proposing mechanism (Theorem
\ref{thm:two-retention-buyer}). 
For the random-offerer mechanism,
we take advantage of the distribution-free IPV screening results of
Bei, Wu, and Zhou \cite{BWZ26} and, with substantially weaker
regularity conditions for $(x, y)$, establish a similar worst-case
bound (Theorem \ref{thm:two-retention-random}).\footnote{Specifically, if $x$ and $y$ are independent, no
  further assumptions are needed. If $x$ and $y$ are dependent,
  log-concavity of the marginal distributions is sufficient.}

%% file: related_work.tex
\subsection{Related Work}

The model and results of this paper can be related to several
well-developed threads of research in economics and computer science.
At a high level, our results differ from most previous work in
economics on bilateral trade with interdependent values due to its focus
on worst-case approximation bounds (exceptions are detailed below).
Our results differ from most previous work in computer science on
approximation guarantees for bilateral trade mechanisms via its
combined focus on 
interdependent values,
simple and widely used mechanisms, 
the objective of maximizing gains from trade,
and equilibrium analysis in dynamic games (again, with exceptions
detailed below).


\vspace{-.5\baselineskip}
\paragraph{Approximate mechanism design with interdependent values.}
One obviously relevant line of work is algorithmic and approximate mechanism
design. Works in this literature typically focus on designing (possibly
complex) mechanisms for which truthful revelation is a
(dominant-strategy or Bayesian-Nash) equilibrium that is guaranteed to
approximately optimize an objective function such as welfare, revenue,
consumer surplus, or gains from trade. 
Early work in this area (beginning with Nisan and Ronen~\cite{NR99})
focused on private-value settings, but a number of more recent papers
consider interdependent values. For example,
Roughgarden and Talgam-Cohen~\cite{RT16} and Li~\cite{L17} identify
conditions (on
valuations and the information structure) under which
prior-independent and near-optimal (for revenue) ex post incentive
compatible and individually rational mechanisms can be derived using
an analog of Myerson's~\cite{M81} theory of virtual values; follow-up
works include Chawla et al.~\cite{CFK14}, Eden et al.~\cite{E+18},
and Eden et al.~\cite{E+21}.
Daskalakis, Papadimitriou, and Tzamos~\cite{DPT16} consider a
specialized interdependent setup with an informed seller similar to that in our
Examples~\ref{ex:pooling} and~\ref{ex:rare}, and optimize expected
revenue over all choices of multi-round mechanisms and seller information
disclosure policies.
Eden et al.~\cite{E+19}, in the context of combinatorial auctions,
introduce the SOS (submodularity over signals) condition on
interdependent valuations, and derive truthful auctions for general
single-parameter settings that guarantee
a constant fraction of the maximum-possible welfare. 
This paper considers publicly known valuation functions (in
Examples~\ref{ex:pooling} and~\ref{ex:rare}, the analog would be
publicly known values for~$a_S$ and~$a_B$) but private signals.
Eden, Goldner, and Zheng~\cite{E+22}
and Eden et al.~\cite{E+23} extend the model of~\cite{E+19} to include
private valuation functions, and (under SOS) obtain polylogarithmic
and constant-factor approximations to the optimal welfare,
respectively.
Dobzinski et al.~\cite{D+25}, in the context of bilateral trade,
systematically study the approximate
welfare achievable by Bayesian incentive-compatible and
interim-individually rational budget-balanced mechanisms, providing
both possibility and impossibility results (depending on agents'
valuation functions).
Dobzinski et al.~\cite{D+25} explicitly mention approximation
guarantees for gains from trade with interdependent values as an open
research direction.

Our work differs from the literature above through its focus on simple
mechanisms (and the appropriate equilibrium concepts to study them)
the gains from trade objective function (as opposed to welfare or
revenue), and guarantees for worst-case equilibria (as opposed to for
a canonical equilibrium). There is also a sequence of results
on approximation guarantees for simple mechanisms with respect to the
welfare objective for the simpler scenario of independent or correlated
private values.\footnote{Note that approximation guarantees with respect to
the gains from trade objective carry over to the welfare objective, but
not vice versa.} With independent private values, Blumrosen and
Dobzinski~\cite{BD14} proved that a fixed-price mechanism captures at
least a constant fraction of the expected welfare. The constant factor
was later improved by~\cite{BD21,KPV22,CW23,LRW23,GdK26}.
Dobzinski and Shaulker~\cite{DS24} show that, with correlated private
values, the buyer-proposing mechanism achieves a
$(1-\tfrac{1}{e})$-approximation of the optimal expected welfare.

\vspace{-.5\baselineskip}
\paragraph{Approximate gains from trade with private values.}
Several works have achieved positive results for approximating
gains from trade with simple mechanisms in private-value settings
(i.e., without the signaling or adverse selection complications of
the interdependent setting).
McAfee~\cite{M08} shows that, when the median of buyer's valuation
distribution is at least that of the seller's valuation distribution,
a fixed price can capture at least half of first-best gains from
trade; Kang and Vondrak~\cite{KV19} characterize the optimal fixed
price in this setting and Colini-Baldeschi et al.~\cite{C+17} generalize
McAfee's result in several different directions.
Blumrosen and Mizarhi~\cite{BM16} show that, when the buyer's
valuation distribution satisfies the monotone hazard rate condition,
the seller-proposer mechanism achieves at least a $1/e$ fraction of
the first-best gains from trade; our Lemma~\ref{lem:screening} is
similar to a key step in their analysis.
Brustle et al.~\cite{B+17} prove that, without any restrictions on the
valuation distributions (other than finite first moments), the
random-proposer mechanism (see Section~\ref{s:mechanisms}) guarantees
at least half of the second-best gains from trade (i.e., the maximum
expected gains from trade achievable by any Bayesian
incentive-compatible, interim individually rational, and ex ante
weakly budget-balanced mechanism).
Babaioff et al.~\cite{B+18} work in the same model as~\cite{B+17} and
prove asympototic full efficiency in large markets.
Deng et al.~\cite{D+22} show that the random-proposer mechanism
obtains a constant fraction of the stronger first-best benchmark.
Fei~\cite{F22}, Jo~\cite{J26}, and Bei, Wu, and
Zhou~\cite{BWZ26} improve the constant factor, culminating in a
bound of~$1/e$.
Fe~\cite{F22} also shows that,
assuming that the buyer's valuation distribution satisfies the monotone
hazard rate condition, the seller-proposer mechanism guarantees at
least a $1/(e-1)$ fraction of the first-best benchmark (and that this
analysis is tight). 
Babaioff et al.~\cite{B+26} consider a much more general matching
market scenario (with arbitrary downward-closed feasibility
constraints)---inspired in part by Cai et al.~\cite{C+21}'s work on
approximate gains from trade in such markets---and shows that an
appropriate generalization of the random-proposer mechanism again
achieves a constant fraction of the full gains from trade.
Bei, Wu, and Zhou~\cite{BWZ26} provide a tight analysis of this
mechanism, establishing an approximation factor of~$1/e$ relative to
the first-best benchmark.

\vspace{-.5\baselineskip}
\paragraph{The price of anarchy in dynamic games.}
While most of the price-or-anarchy literature concerns static
games, a small consider dynamic games.
For example, Paes Leme, Syrgkanis, and Tardos~\cite{PST12}
prove bounds on the price of anarchy of subgame-perfect equilibria 
in complete-information sequential first-price item
auctions with unit-demand bidders.
Syrgkanis and Tardos~\cite{ST12} study the price of anarchy of
sequential auctions in more general settings (e.g., matching markets);
their positive results apply generally to Bayesian-Nash equilibria of
their games and do not require sequential rationality.
Feldman, Lucier, and Syrgkanis~\cite{FLS13} consider sequential
first-price auctions with more general classes of bidder valuations.
Babaioff et al.~\cite{B+23} consider perfect Bayesian equilibria (PBE)
of a two-stage game that models auctions followed by a resale market
and study which first-stage auction formats leads to good welfare
guarantees; like~\cite{ST12}, the positive results hold for all
Bayesian-Nash equilibria and do not require sequential rationality.
More distantly related is recent work by Banerjee et al.~\cite{B+26b}
on the price of anarchy in a competitive version of the Bayesian
persuasion framework of Kamenica and Gentzkow~\cite{KG11}; in that
framework, unlike ours, the sender (analogous to our proposer) has ex
ante commitment power for their information disclosure policy.

\vspace{-.5\baselineskip}
\paragraph{Markets for lemons.}
Akerlof~\cite{akerlof1970} shows how adverse selection with
  asymmtric information can lead to no trade at equilibrium, even in
  markets in which positive gains from trade are guaranteed.
Akerlof's rational expectations analysis does not consider any
  specific mechanism or price formation
  process. Wilson~\cite{wilson80} considers three mechanisms for
  price-setting (seller-proposing, buyer-proposing, and
  auctioneer-proposing) and provides a partial characterization of
  equilibria in each case.
Levin~\cite{levin01} studies the effect of varying the amount of
seller information on the probability of exchange.
Voorneveld and Weibull~\cite{VW11} consider the seller-proposer
mechanism with uncertain item equality and an informed seller (similar
to our Examples~\ref{ex:pooling} and~\ref{ex:rare}) and characterize
the structure of perfect Bayesian equilibria. They examine gains from
trade at equilibrium---e.g., when it is fully realized, how they are
split between the buyer and seller---but not from the worst-case
approximation viewpoint adopted here.
Kartik and Zhong~\cite{KZ26} work with a similar setup and characterize
on of the set of attainable equilibrium payoffs as the information
structure varies. One specific consequence of their work is that,
without some type of restriction on the information structure (as we
do in this paper), no meaningful worst-case equilibrium approximation
of gains from trade is possible; the same conclusion also follows from
the impossibility result in Brooks and Du~\cite{BD23}.
The work of Abraham, Athey, and Babaioff~\cite{AAB20} on the analysis
of common-value second-price auctiuons with asymmetrically informed
bidders is only distantly related, although pessimistic signals do
play an interesting role in their seller revenue analysis (the
expected item value conditioned on such a signal, rather than the
likelihood of such a signal, which is the relevant parameter in the
present work).

%% file: model.tex
\section{Model}\label{sec:model}

\subsection{Types, Valuations, and Signals}

We consider a model in which there is a seller who holds an asset, and
a buyer who may wish to purchase it. The buyer and seller each have a
private type $\theta_B, \theta_S \in \Theta$, where $\Theta$ is a
continuous type space, such as $\R$. These are drawn independently and
identically according to a distribution $\mc G$ over $\Theta$. There
is also a common component and signal
$(C, \Sigma_S, \Sigma_B) \in \R \times \mc S_S \times \mc S_B$ that is
drawn independently from the types $\theta_B, \theta_S$. The common
component has marginal distribution $\mc F$. The signals
$(\Sigma_S, \Sigma_B)$ may each indicate arbitrary information
regarding the common component, except for the restriction that the
signal spaces have finite cardinality
$\abs{\mc S_S}, \abs{\mc S_B} < \infty$. Given a type $\theta$ and a
realization $c$ of the common component, the buyer or seller has a
willingness to pay of $v(c, \theta)$. In this paper, we exclusively
consider $v$ that are affine in the common component with nonnegative
slope:
\begin{equation}\label{eq:affine}
    v(c,\theta)=c\,x(\theta)+y(\theta),
    \qquad x(\theta)\ge 0.
\end{equation}
where $x$ and $y$ are measurable functions. The definitions of
$(v, \mc G, \mc F, \Sigma_S, \Sigma_B)$ are common knowledge to both
the buyer and the seller. We call such a tuple a \textit{market
  instance}. It is further common knowledge that the seller observes
the realization of the signal $\Sigma_S$ and the buyer observes the
realization of $\Sigma_B$. Both parties observe the realization of
their own type.

\vspace{-.1in}
\paragraph{Affine valuations: applications.}
Affine valuations cover a broad range of familiar sources of
heterogeneity in interdependent valuations. Purely multiplicative
valuations are versatile in their own right. They appear, for example,
in standard market microstructure when the asset being exchanged is a
security \cite{glostenmilgrom1985}. Here, $C$ can be interpreted as
the uncertain future cash flow to which the seller holds title. The
value of $x(\theta)$ can be interpreted as an idiosyncratic discount
factor, reflecting the trader's private urgency for
liquidity. Alternatively, the common component $C$ can be interpreted
as a quality parameter for the asset, multiplied by an idiosyncratic
willingness to pay for quality \cite{akerlof1970,mussarosen1978}. The
additive term $y(\theta)$ may further model private holding costs
\cite{duffiegarleanupedersen2005}. It can also capture
inventory-dependent adjustments to the value of a fixed-size trade,
motivated by mean--variance models with heterogeneous initial holdings
and quasilinear cash payments \cite{rostekyoon2021}.

\vspace{-.1in}
\paragraph{Restrictions on signals.}
To model the case where only one party is informed, such as in the
earlier examples where only the seller was informed, one could simply
take the uninformed party's signal to be a constant
$\Sigma_B = \sigma_B \in \mc S_B$, so that it provides no information
about the common component. Buyer and seller signals are allowed to be
correlated, even conditional on the realization of $C$. We do require
that there is a consistent interpretation of on party's signal as
``good news'' or ``bad news'' regardless of the realization of the
other party's signal. Define, for $\sigma_S \in \mc S_S$ and
$\sigma_B \in \mc S_B$,
\[
    m(\sigma_S, \sigma_B)
    :=
    \ex{C \mid \Sigma_S = \sigma_S, \Sigma_B = \sigma_B}
\]
to be the expected value of the common component given the signals of both parties. We require that for any $\sigma_S, \sigma'_S \in \mc S_S$ and $\sigma_B, \sigma'_B \in \mc S_B$,
\begin{align}
    m(\sigma_S, \sigma_B) \le m(\sigma_S, \sigma_B')
    &\implies
    m(\sigma_S', \sigma_B) \le m(\sigma_S', \sigma_B')\\
    m(\sigma_S, \sigma_B) \le m(\sigma_S', \sigma_B)
    &\implies
    m(\sigma_S, \sigma_B') \le m(\sigma_S', \sigma_B').
\end{align}
When only one party is informed (e.g. with only an informed seller), this condition holds trivially. For expositional simplicity, we assume that every signal pair $(\sigma_S, \sigma_B) \in \mc S_S \times \mc S_B$
is realized with positive probability.\footnote{The IPV reduction extends
to incomplete support under common signal orderings: one may
pair the most pessimistic seller signal with the most optimistic
buyer signal compatible with it, or, symmetrically, the most
optimistic buyer signal with the most pessimistic seller signal
compatible with it. The reduction then uses the forecast and
probability of the selected pair. Absolute-constant guarantees
additionally require the corresponding rarity bounds.}

\subsection{Actions and Payoffs}

In this subsection, we formally define a two-stage proposer game, corresponding to the seller- and buyer-proposing mechanisms. We generally consider a setting in which there a \textit{proposer} $P$ and a \textit{receiver} $R$, where $(P, R) \in \set{(S, B), (B, S)}$. The game proceeds in two stages. First, the proposer, after observing her type $\theta_P$ and signal realization $\sigma_P$, may choose to list a quote within a finite menu $Q \subset \R$ of prices, or choose to exit (denoted by $\bot$). Formally, letting $\mc Q := Q \cup \set{\bot}$, the proposer selects a (potentially mixed) proposal strategy
\[
    p: \Theta \times \mc S_P \to \Delta(\mc Q)
\]
where $\Delta(\mc Q)$ denotes probability distributions over $\mc Q$. We write $p_q(\theta,\sigma_P)$ for the probability assigned to quote $q\in\mc Q$ when the proposer has type $\theta$ and signal $\sigma_P$. Let $\widetilde q$ denote the realized quote, drawn using the proposer's private randomization according to
\[
    \widetilde q\mid(\theta_P,\Sigma_P)
    \sim p(\theta_P,\Sigma_P).
\]
The receiver then observes the proposal before selecting to accept or reject the quote. Formally, the receiver selects an acceptance strategy that takes as input their type, signal realization, and the proposer's quote, and selects a probability of accepting the offer:
\[
    a: \Theta \times \mc S_R \times \mc Q \to [0,1].
\]
We set $a(\theta,\sigma_R,\bot)=0$, since exit precludes trade.

If the proposer is a seller and the receiver is a buyer, then the quote is an asking price for the asset. If the proposer is a buyer and the receiver is a seller, then the quote is a bid price. Because we will use gains from trade as a benchmark, with first-best gains from trade given by
\begin{equation}
    \mathrm{FB}
    :=
    \ex{\pa{v(C, \theta_B) - v(C, \theta_S)}_+},
\end{equation}
we normalize the seller's utility by subtracting their utility from holding the asset in the case that trade does not take place. For a cash quote $q\in Q$, the buyer's and seller's payoffs, averaged over the receiver's randomization, are given by
\begin{align}
    u_B(q,a)
    &:=
    a(\theta_R,\Sigma_R,q)
    \bra{v(C,\theta_B)-q},\\
    u_S(q,a)
    &:=
    a(\theta_R,\Sigma_R,q)
    \bra{q-v(C,\theta_S)}.
\end{align}
We set $u_B(\bot,a)=u_S(\bot,a)=0$. Under proposal strategy $p$, the resulting payoffs are $u_B(\widetilde q,a)$ and $u_S(\widetilde q,a)$, where expectations also average over the proposer's randomization.

The two-stage proposer game is a signaling game. The proposer's quote serves two roles: it simultaneously sets the terms of exchange while also providing some information to the receiver about what the proposer knows about the common component's value given $\Sigma_P$. The proposer must also be wary of adverse selection, as acceptance by the receiver also indicates information pertaining to the common component's value given $\Sigma_R$.

\subsection{Sequential Equilibria}\label{ss:eq}

\paragraph{The subtlety: the receiver's interpretation of off-path prices.}
We would like to define our game as a two-stage extensive form game
with subgame perfection. This would require sequential rationality on
the part of the receiver---intuitively, that the receiver always best
responds in the second-stage subgame that is actually realized
(including in ``off-path'' subgames that would not be reached should
the proposer follow their equilibrium strategy).
As is typical in
signaling games, there is some complexity in making this
well defined. The proposer's quote affects the receiver's valuation
for the asset, as it communicates information about the proposer's
signal. But what information does it communicate? 
With respect to a fixed equilibrium,
the receiver could apply Bayes' rule to form a posterior over the
asset's value given the set of proposers who choose to set a
particular quote. While Bayes' rule resolves the receiver's inference
on-path, what does the receiver infer about the asset's value at
prices that are not realized on the equilibrium path? The proposer's
strategy may not involve using the full set of quotes $\mc Q$ (cf.,
Examples~\ref{ex:pooling} and~\ref{ex:rare}). Because off-path prices
are quoted with zero probability, Bayes' rule cannot be used. However,
this inference must be well-defined so that the proposer's
best-response problem remains well-defined. The off-path
interpretation of a quote by the receiver affects the proposer's
choice of what to play on-path.

\vspace{-.1in}
\paragraph{Supplementing strategies with beliefs.}
We handle this in the usual way by augmenting the equilibrium to
explicitly consider the beliefs $\lambda$ of the receiver in addition
to the strategies $(p, a)$ played by the proposer and
receiver. Formally, we consider triples of the form $(p, a, \lambda)$,
where $p$ and $a$ are valid strategies for
the proposer and receiver, and
$\lambda: \mc S_R \times Q \to \Delta(\mc S_P)$ denotes the receiver's
beliefs over the proposer's signal given their own signal and the
proposer's quote (excluding proposal exit $\bot$). Receiver beliefs
$\lambda$ allow us to compute a well-defined expectation for any
function $h$ of the common component $C$ given the receiver's signal
under their interpretation of a proposer quote $q \in Q$:
\[
    \exx{\lambda}{
        h(C)\mid\widetilde q=q,\Sigma_R=\sigma_R
    }
    :=
    \sum_{\sigma_P\in\mc S_P}
    \lambda_{\sigma_P}(\sigma_R,q)
    \ex{
        h(C)\mid
        \Sigma_P=\sigma_P,\Sigma_R=\sigma_R
    }.
\]

\begin{definition}[Sequential Equilibrium]
A triple $(p, a, \lambda)$ is a {\em sequential equilibrium} if the
following conditions hold:
\begin{enumerate}
    \item \textit{Receiver best response: }At each receiver signal $\sigma_R\in\mc S_R$,
    every cash quote $q\in Q$, and for $\mc G$-almost
    every $\theta\in\Theta$, the receiver accepts with
    probability one if
    \[
        \begin{cases}
            \exx{\lambda}{
                v(C,\theta)\mid
                \widetilde q=q,\Sigma_R=\sigma_R
            }-q,
            & R=B,\\[3pt]
            q-\exx{\lambda}{
                v(C,\theta)\mid
                \widetilde q=q,\Sigma_R=\sigma_R
            },
            & R=S
        \end{cases}
    \]
    is strictly positive, and rejects with probability
    one if it is strictly negative. If it is zero,
    any acceptance probability is permitted.

    \item \textit{Proposer best response:} At each proposer signal $\sigma_P\in\mc S_P$,
    for $\mc G$-almost every $\theta\in\Theta$, and
    every $q\in\mc Q$,
    \begin{equation}
        p_q(\theta,\sigma_P)>0
        \implies
        q\in
        \argmax_{q'\in\mc Q}
        \ex{
            u_P(q',a)
            \mid
            \Sigma_P=\sigma_P,\theta_P=\theta
        }.
    \end{equation}

    \item \textit{On-path belief consistency:} The beliefs $\lambda$
      are consistent with Bayes' rule on-path. That is, 
for every quote $q$ reached with positive probability conditioned 
on the receiver's signal $\sigma_R \in \mc S_R$, 
    \begin{equation}
        \lambda_{\sigma_P}(\sigma_R,q)
        =
        \frac{
            \ell_q(\sigma_P)
            \pr{\Sigma_P=\sigma_P\mid\Sigma_R=\sigma_R}
        }{
            \sum_{\sigma_P'\in\mc S_P}
y            \ell_q(\sigma_P')
            \pr{\Sigma_P=\sigma_P'\mid\Sigma_R=\sigma_R}
        },
    \end{equation}
where
    \begin{equation}
        \ell_q(\sigma_P)
        :=
        \int p_q(\theta,\sigma_P)\,d\mc G(\theta)
    \end{equation}
denotes the probability of quote $q \in \mc Q$ being used by the
    proposer conditioned the proposer observing $\sigma_P \in \mc
    S_P$.\footnote{The likelihood $\ell$ is common across receiver signals as the proposer does not observe the receiver's signal $\sigma_R$, and therefore cannot signal in a way that depends on it.}

    \item \textit{Off-path belief consistency.} 
Off-path beliefs must be consistent with a common sequence of
perturbed quote likelihoods. Specifically, there exists a sequence
of likelihoods over proposer actions $(\ell^n)_{n\ge 1}$ such that
\[
    \ell_q^n(\sigma_P)>0,
    \qquad
    \sum_{q\in\mc Q}\ell_q^n(\sigma_P)=1,
    \qquad
    \ell_q^n(\sigma_P)\longrightarrow\ell_q(\sigma_P)
\]
for every proposer signal $\sigma_P$ and quote $q\in\mc Q$,
with the first two conditions holding for every $n$.
For every receiver signal $\sigma_R$ and cash quote $q\in Q$
that is reached with probability zero conditional on $\sigma_R$,
we require
\begin{equation}\label{eq:offpath}
    \lambda_{\sigma_P}(\sigma_R,q)
    =
    \lim_{n\to\infty}
    \frac{
        \ell_q^n(\sigma_P)
        \cdot
        \pr{\Sigma_P=\sigma_P\mid\Sigma_R=\sigma_R}
    }{
        \sum_{\sigma_P'}
        \ell_q^n(\sigma_P')
        \cdot
        \pr{\Sigma_P=\sigma_P'\mid\Sigma_R=\sigma_R}
    }
    \qquad\text{for every }\sigma_P.
\end{equation}
Thus, all off-path beliefs must arise from Bayes' rule along
the same sequence of perturbed likelihoods, common across
receiver signals. Strict positivity ensures that these
Bayesian updates are well-defined at every step.
\end{enumerate}
\end{definition}

Our equilibrium concept corresponds to 
the sequential equilibrium of
Kreps and Wilson \cite{krepswilson1982},
adapted for our
continuous type space $\Theta$. At a high level, sequential equilibria
explain off-path prices in terms of beliefs the receiver would have
had from Bayes' rule given vanishing full-support trembles of the
proposer's strategy. Because the proposer's type $\theta_P$ is
independent of $(C, \Sigma_P, \Sigma_R, \theta_R$), we can further
omit inferences regarding $\theta_P$ from the receiver's
beliefs. Sequential equilibria are a refinement of weak Perfect
Bayesian Equilibria, which make no restriction on off-path beliefs.

\begin{theorem}\label{thm:eqexist}
    For any market instance
    $(v,\mc G,\mc F,\Sigma_S,\Sigma_B)$,
    any finite price menu $\mc Q$ including proposer
    exit $\bot$, and either choice of proposer
    $P\in\{S,B\}$, there exists a sequential
    equilibrium $(p,a,\lambda)$.
\end{theorem}

\paragraph{Beliefs in the introductory examples.}
For completeness, we overview the corresponding beliefs $\lambda$ in the equilibria of Examples \ref{ex:pooling} and \ref{ex:rare}. In Example~\ref{ex:pooling}, the buyer assigns probability
$2/3$ to low quality at prices on the lower branch, and
probability one to high quality at the pooled price and
on the upper branch. Every unused price is interpreted
as indicating low quality. In Example~\ref{ex:rare},
every cash offer, whether used or unused, is interpreted
as coming from a low-signal seller, yielding an expected
quality of $t/2$ (for $t>0$).
The off-path beliefs for both examples can be understood through vanishing
trembles: sellers usually follow their prescribed strategies,
but occasionally choose an action from a common full-support
distribution. Let trembles after high news become negligible
relative to trembles after low news as both probabilities
vanish. Since unused quotes arise only from trembles,
observing one leads the buyer to assign probability approaching
one to low news. At used quotes, the limiting beliefs instead
recover the Bayesian beliefs induced by the prescribed strategy. This sequence of trembles induce the likelihoods $\ell^n$ in \eqref{eq:offpath}.

\subsection{Equilibrium Gains from Trade}

For each proposer $P\in\{S,B\}$, let
$\mc E_P(\mc I,\mc Q)$ denote the set of all sequential
equilibria of the corresponding proposer game given
market instance $\mc I$ and finite price menu $\mc Q$.
In general, there may be vast equilibrium multiplicity
for a fixed instance, proposer, and price menu.

The expected
gains from trade at an equilibrium
$(p,a,\lambda)\in\mc E_P(\mc I,\mc Q)$, as a fraction
of first-best, are given by
\begin{equation}
    W_P(p,a,\lambda)
    :=
    \frac{1}{\mathrm{FB}}
    \ex{
        \sum_{q\in Q}
        p_q(\theta_P,\Sigma_P)
        a(\theta_R,\Sigma_R,q)
        \pa{v(C,\theta_B)-v(C,\theta_S)}
    }.
\end{equation}
The sum averages over the proposer's randomization,
while $a(\theta_R,\Sigma_R,q)$ accounts for the
receiver's randomization. Exit contributes zero
gains from trade and is therefore omitted from the sum.

In this paper, we focus on characterizing the
worst-case fraction of first-best:
\begin{equation}
    W^-_P(\mc I,\mc Q)
    :=
    \inf_{(p,a,\lambda)\in\mc E_P(\mc I,\mc Q)}
    W_P(p,a,\lambda).
\end{equation}
To avoid dependence on the price menu, we take the
limit as the price menu becomes fine. Formally, we
call a sequence of finite menus
$(\mc Q_n)$, where $\mc Q_n=Q_n\cup\{\bot\}$, fine if
\begin{equation}\label{eq:menus}
    \sup_{|q|\le H}\operatorname{dist}(q,Q_n)
    \longrightarrow0
    \quad\text{for every }H>0,
    \qquad
    \operatorname{dist}(q,Q)
    :=
    \inf_{q'\in Q}|q-q'|.
\end{equation}
We then define
\begin{equation}
    W^-_P(\mc I)
    :=
    \inf_{\text{fine menus }(\mc Q_n)}
    \liminf_{n\to\infty}
    W^-_P(\mc I,\mc Q_n).
\end{equation}

%% file: reduction.tex
\section{Reduction to an IPV market} \label{sec:reduction}

We begin by proving a lower bound on equilibrium gains from trade in the special case where there is an informed seller who is proposing to an uninformed buyer. Our high-level proof strategy proceeds in two main steps. We first construct an auxiliary market where buyer and seller values are independent. Under suitable regularity conditions on buyer and seller valuations in the original market, we show in this section that equilibrium gains from trade must exceed a fraction of the first-best gains from trade in this IPV market. Then, in Section \ref{sec:retention}, we show that under some additional regularity, the IPV market's first-best lower bounds the original market's first-best. Together, this produces a lower bound on the fraction of first-best attained at any equilibrium.

\subsection{Belief-Robust Values}

As is common in quantifications of inefficiency at equilibrium, we lower bound equilibrium gains from trade by considering deviations by the seller that guarantee a certain amount of utility. Best-response by the seller then guarantees that at any equilibrium, seller utility must exceed that of the deviation. We specifically consider deviations that are robust to the beliefs held by the buyer. As it turns out, we can understand these deviations in terms of the seller's profit that would be obtained in an IPV market where both the buyer and seller publicly observe the most pessimistic seller signal.

Formally, define
\begin{equation}
    \sigma^-_S \in \argmin_{\sigma_S \in \mc S_S}
    \ex{C \mid \Sigma_S = \sigma_S},
    \qquad
    d := \ex{C \mid \Sigma_S = \sigma_S^-}
\end{equation}
to be the most pessimistic seller signal (or one of them, if there are multiple), and the corresponding most pessimistic common component forecast. We can then define the IPV market that arises if both the seller and buyer were to publicly observe $\sigma_S^-$ as follows:
\begin{equation}
    \hat v_S := dx(\theta_S) + y(\theta_S),
    \qquad
    \hat v_B := dx(\theta_B) + y(\theta_B)
\end{equation}
Notice that these valuations are now independent. First-best gains from trade in this IPV market are then given by
\begin{equation}
    R := \ex{\pa{\hat v_B - \hat v_S}_+}
\end{equation}
We call this market the \textit{belief-robust} market. We are now ready to formally state the main result of this section: that equilibrium gains from trade is at least a constant fraction of the belief-robust market's first-best gains from trade.

\begin{theorem}\label{thm:reduction}
    Let $\mc I := (v,\mc G,\mc F,\Sigma_S,\Sigma_B)$ be an
    instance in which the seller proposes and the buyer is
    uninformed. Suppose that $\ex{|C|}<\infty$,
    $\mathrm{FB}>0$, and, under $\mc G$, $x(\theta)$ and
    $y(\theta)$ have finite first moments and a jointly
    log-concave density. Then
    \begin{equation}
        W^-_S(\mc I)
        \ge
        \frac{\pr{\Sigma_S=\sigma_S^-}}{e}
        \cdot\frac{R}{\mathrm{FB}}.
    \end{equation}
\end{theorem}

\begin{proof}
There are two main steps in this proof. In the first step,
we show that a seller who observes $\sigma_S^-$ may always
secure at least her monopoly profit in the belief-robust
market. We then utilize log-concavity to show that this
profit captures at least a $\frac1e$ fraction of $R$.
The factor $\pr{\Sigma_S=\sigma_S^-}$ arises because we
only count the profits of sellers who observe $\sigma_S^-$;
all other sellers obtain nonnegative expected utility
because they may exit.

Suppose we have a finite price menu
$\mc Q=Q\cup\{\bot\}$. For a seller with reservation value
$k\in\R$, define
\begin{equation}\label{eq:grid-profit}
    u_{\mc Q}(k)
    :=
    \max\left(
        \{0\}\cup
        \left\{
            (q-k)\pr{\hat v_B>q}:
            q\in Q,\ q>k
        \right\}
    \right).
\end{equation}
This gives the monopoly profit that the seller obtains
in the belief-robust market when restricted to the
price menu $\mc Q$. Notice that we have required strict
acceptance by the buyer, so our lower bound will not
depend on how the buyer breaks ties.

\begin{lemma}\label{lem:sellerprofit}
    Every equilibrium
    $(p,a,\lambda)\in\mc E_S(\mc I,\mc Q)$
    of the seller-proposing game satisfies
    \begin{equation}\label{eq:finite-sellerprofit}
        W_S(p,a,\lambda)\,\mathrm{FB}
        \ge
        \pr{\Sigma_S=\sigma_S^-}
        \cdot\ex{u_{\mc Q}(\hat v_S)}.
    \end{equation}
\end{lemma}

\begin{proof}
Fix such an equilibrium $(p,a,\lambda)$.
Notice first that, at every
on- or off-path price $q$, the buyer's forecast is
a mixture of the seller's signal-conditioned
forecasts. Each of these forecasts is at least $d$,
and hence
\[
    \exx{\lambda}{
        C\mid\widetilde q=q
    }\ge d.
\]
Since $x(\theta_B)\ge0$, a higher forecast of the
common component cannot decrease the buyer's
valuation
\[
    \exx{\lambda}{
        C\mid\widetilde q=q
    }x(\theta_B)+y(\theta_B)
    \ge
    dx(\theta_B)+y(\theta_B)
    =
    \hat v_B.
\]
The buyer's expected valuation at the quote is
therefore at least her valuation in the
belief-robust market. As such, any buyer with
$\hat v_B>q$ strictly prefers to accept at that
price under any permissible beliefs, including
$\lambda$.

Now consider a seller with signal $\sigma_S^-$ and
type $\theta$, and write $k=dx(\theta)+y(\theta)$.
If this seller quotes an available price $q>k$,
the probability of acceptance is at least
$\pr{\hat v_B>q}$.

Because the buyer is uninformed, acceptance of this
fixed quote depends only on the buyer's type and
randomization. These are independent of the common
component and the seller's signal. Consequently,
the seller's expected profit from quoting $q$ is
\begin{align*}
    &\ex{
        \bigl(q-v(C,\theta)\bigr)
        a(\theta_B,\Sigma_B,q)
        \mid \Sigma_S=\sigma_S^-
    }\\
    &\qquad=
    (q-k)\ex{a(\theta_B,\Sigma_B,q)}
    \ge
    (q-k)\pr{\hat v_B>q}.
\end{align*}
Because the seller may also exit, equilibrium
best response guarantees expected utility of at least
$u_{\mc Q}(k)$.

Conditional on $\Sigma_S=\sigma_S^-$, the seller's
type still has distribution $\mc G$. Averaging over
these sellers therefore gives expected profit of at
least $\ex{u_{\mc Q}(\hat v_S)}$. Multiplying by
$\pr{\Sigma_S=\sigma_S^-}$ accounts for the probability
of this signal. All other sellers obtain nonnegative
expected utility because they may exit.

Finally, the buyer's ex-ante expected utility is
nonnegative by best response and on-path Bayesian
updating. Gains from trade equal the sum of the
traders' utilities, and hence are at least seller
profit. This proves \eqref{eq:finite-sellerprofit}.
\end{proof}

We can now define the continuous analogue of
$u_{\mc Q}$ as
\begin{equation}\label{eq:continuous-profit}
    u(k)
    :=
    \sup_{q>k}(q-k)\pr{\hat v_B>q}.
\end{equation}
Before applying a screening bound to $u$, we show
that this is indeed the limit of the finite-menu
problem. We do so in the following Lemma.

\begin{lemma}\label{lem:finemenu}
    For any integrable buyer value $\hat v_B$ and
    any fine menu sequence $(\mc Q_n)$,
    \[
        \lim_{n\to\infty}u_{\mc Q_n}(k)=u(k)
        \qquad\text{for every }k\in\R.
    \]
    Moreover, for any integrable seller value
    $\hat v_S$,
    \begin{equation}\label{eq:finemenu-expectation}
        \lim_{n\to\infty}
        \ex{u_{\mc Q_n}(\hat v_S)}
        =
        \ex{u(\hat v_S)}.
    \end{equation}
\end{lemma}

\begin{proof}
Fix $k\in\R$ and $q>k$. Because the menus become
fine, we can choose prices $q_n\in Q_n$ with
$q_n<q$ and $q_n\to q$, by approximating prices
just below $q$. Eventually $q_n>k$, and
$\pr{\hat v_B>q_n}\ge\pr{\hat v_B>q}$. Thus,
\[
    \liminf_{n\to\infty}u_{\mc Q_n}(k)
    \ge
    (q-k)\pr{\hat v_B>q}.
\]
Taking the supremum over $q>k$ gives
$\liminf_n u_{\mc Q_n}(k)\ge u(k)$.
The reverse inequality holds for every $n$,
since restricting the available prices cannot
increase monopoly profit. This proves pointwise
convergence.

To pass to expectations, notice that
\[
    0\le u_{\mc Q_n}(k)\le u(k)
    \le \ex{(\hat v_B-k)_+}
    \le \ex{|\hat v_B|}+|k|.
\]
Since $\hat v_S$ is integrable, dominated
convergence gives \eqref{eq:finemenu-expectation}.
\end{proof}

It remains to bound monopoly profit in the
belief-robust market. We use Gui and Huang's
formulation of the Condorelli inequality
\cite[Theorem~1.1]{guihuang2022} to obtain the
following screening bound.

\begin{lemma}\label{lem:screening}
    If $\hat v_B$ is integrable and has a
    log-concave survival function, then for every
    $k\in\R$,
    \begin{equation}\label{eq:screening}
        u(k)\ge
        \frac1e\ex{(\hat v_B-k)_+}.
    \end{equation}
\end{lemma}

\begin{proof}
For every fixed $k$, the function
$t\mapsto\pr{\hat v_B>k+t}$ is nonnegative,
measurable, and log-concave on $[0,\infty)$.
Gui and Huang's theorem therefore yields
\begin{align*}
    \ex{(\hat v_B-k)_+}
    &=
    \int_0^\infty \pr{\hat v_B>k+t}\,dt\\
    &\le
    e\sup_{t>0}t\pr{\hat v_B>k+t}\\
    &=e\,u(k),
\end{align*}
where the last equality substitutes $q=k+t$.
Rearranging gives the desired result.
\end{proof}

Putting everything together to conclude the proof, begin by taking any fine menu sequence $(\mc Q_n)$.
Lemma~\ref{lem:sellerprofit} holds for every
equilibrium on each menu. Taking the infimum over
equilibria, then the lower limit over menus, and
applying Lemma~\ref{lem:finemenu}, gives
\begin{align}
    \mathrm{FB}\cdot
    \liminf_{n\to\infty}W^-_S(\mc I,\mc Q_n)
    &\ge
    \pr{\Sigma_S=\sigma_S^-}
    \cdot
    \lim_{n\to\infty}
    \ex{u_{\mc Q_n}(\hat v_S)}
    \notag\\
    &=
    \pr{\Sigma_S=\sigma_S^-}
    \cdot\ex{u(\hat v_S)}.
    \label{eq:limiting-sellerprofit}
\end{align}

Notice that
$\hat v_B=dx(\theta_B)+y(\theta_B)$ is an affine
transformation of a jointly log-concave random
vector. Its distribution, and therefore its
survival function, are log-concave
\cite[Proposition~3.1 and Theorem~3.3]{saumardwellner2014}.
We may therefore apply Lemma~\ref{lem:screening}.

Because $\hat v_S$ and $\hat v_B$ are independent,
for almost every $k$ under the distribution of
$\hat v_S$,
\[
    u(k)
    \ge
    \frac1e\ex{(\hat v_B-k)_+}
    =
    \frac1e
    \ex{
        (\hat v_B-\hat v_S)_+
        \mid \hat v_S=k
    }.
\]
Averaging over $\hat v_S$ yields
\[
    \ex{u(\hat v_S)}
    \ge
    \frac1e\ex{(\hat v_B-\hat v_S)_+}
    =
    \frac{R}{e}.
\]
Combining this with \eqref{eq:limiting-sellerprofit},
dividing by $\mathrm{FB}$, and taking the infimum
over fine menu sequences proves the theorem.
\end{proof}

%% file: retention.tex
\section{A Lower Bound on Equilibrium Gains From Trade}\label{sec:retention}

In this section, we apply the IPV reduction established in the previous section to show that under some additional regularity, worst equilibrium welfare is at least an absolute constant of first-best gains from trade, so long as the rarity of the most pessimistic forecast is controlled.

Our work in the previous section showed that 
\[W^-_S(\mc I) \ge \frac{\pr{\Sigma_S = \sigma_S^-}}{e} \cdot \frac{R}{\mathrm{FB}}.\]
This bound already goes to zero if $\pr{\Sigma_S = \sigma_S^-}$ can be arbitrarily small. The main contribution of this section is to show that under reasonable regularity conditions, the converse is true as well: if $\pr{\Sigma_S = \sigma_S^-} > 0$, then our bound is also positive. In particular, $R$ is also positive.

For Theorem \ref{thm:reduction} to hold, we already assumed that $x(\theta)$ and $y(\theta)$ have a joint density that is log-concave. We need two additional regularity requirements:
\begin{enumerate}
	\item The distribution $\mc F$ over the common component $C$ is no more dispersed than an exponential random variable with the same mean. This is satisfied, for example, by any $\mc F$ with monotone hazard rate. This condition allows a lower bound on the most pessimistic signal's rarity to induces a lower bound on the expected value of the common component given that signal's realization. 
	\item The quantities $x(\theta)$ and $y(\theta)$ are not perfectly negatively correlated. Perfect negative correlation can cause variation in valuations to cancel out under a pessimistic forecast. This causes $R$ to go to zero (and may cause equilibrium gains from trade to go to zero, as in Example \ref{ex:correlation}) even if the original first-best $\mathrm{FB} > 0$. 
\end{enumerate}

We formalize our full regularity conditions below:

\begin{definition}[$(\pi^-, \rho)$-admissible instance]
	An instance $(v, \mc G, \mc F, \Sigma)$ is $(\pi_-, \rho)$-admissible if 
	\begin{enumerate}
		\item $\mc F$ has non-negative support and is dominated in convex order by an exponential variable of mean $\mu := \exx{\mc F}{C} \in (0, \infty)$. That is, for any convex function $\phi: \R \to \R$, 
		\[\ex{\phi(C)} \le \ex{\phi(Z)}\]
		where $Z$ is an exponential variable with mean $\mu$. 
		\item The rarity of the most pessimistic signal is lower bounded: \[\pr{\Sigma_S = \sigma_S^-} \ge \pi^- > 0\]
		\item For $\theta \sim \mc G$, $(x(\theta), y(\theta))$ has a log-concave density and $\ex{x(\theta) + \abs{y(\theta)}} < \infty$
		\item $\operatorname{Corr}(x(\theta), y(\theta)) \ge \rho > -1$. If either $x(\theta)$ or $y(\theta)$ is constant, then take this to be satisfied for $\rho = 0$. 
	\end{enumerate}
\end{definition}

We are now ready to state the main result of this section. 
\begin{theorem}\label{thm:retention}
    For any $(\pi^-,\rho)$-admissible instance $\mc I$
    in which the seller proposes to an uninformed
    buyer and $\mathrm{FB}>0$,
    \[
        W^-_S(\mc I)
        \ge
        \frac{(\pi^-)^2}{4e}\sqrt{1+\rho}.
    \]
\end{theorem}
\begin{proof}
By Theorem~\ref{thm:reduction}, it suffices to show
that the belief-robust market retains a constant
fraction of the original first-best. Specifically,
we will prove that
\begin{equation}\label{eq:retention-target}
    \frac{R}{\mathrm{FB}}
    \ge
    \frac{\pi^-}{4}\sqrt{1+\rho}.
\end{equation}
There are two main steps in this proof. We first
show that the most pessimistic forecast $d$ cannot
be too small relative to the prior mean $\mu$ by applying the convex-order domination condition.
We then show that the correlation restriction
prevents variation in valuations from cancelling
under this forecast.

We begin with the implication of convex-order
domination that we will need. We state it for an
arbitrary event so that it can also be applied in our two-sided information result in the next section. 

\begin{lemma}\label{lem:forecast}
    Suppose $C\ge0$ is dominated in convex order
    by an exponential random variable with mean
    $\mu=\ex{C}\in(0,\infty)$. Then every event $A$
    satisfies
    \[
        \ex{C\mathbf1_A}
        \ge
        \frac{\mu}{2}\pr{A}^2.
    \]
    Consequently, if $\pr{A}>0$,
    \[
        \ex{C\mid A}
        \ge
        \frac{\mu}{2}\pr{A}.
    \]
\end{lemma}

\begin{proof}
For every $t\ge0$, convex-order domination gives
\[
    \ex{(C-t)_+}\le\mu e^{-t/\mu}.
\]
Using the identity
$(t-C)_+=t-C+(C-t)_+$ and the inequality
$e^{-z}\le1-z+z^2/2$ for $z\ge0$, we obtain
\begin{align*}
    \ex{(t-C)_+}
    &=
    t-\mu+\ex{(C-t)_+}\\
    &\le
    t-\mu+\mu e^{-t/\mu}\\
    &\le
    \frac{t^2}{2\mu}.
\end{align*}
We can therefore bound the expected value of $C$
on the event $A$ as follows:
\begin{align*}
    \ex{C\mathbf1_A}
    &=
    t\pr{A}+\ex{(C-t)\mathbf1_A}\\
    &\ge
    t\pr{A}-\ex{(t-C)_+\mathbf1_A}\\
    &\ge
    t\pr{A}-\ex{(t-C)_+}\\
    &\ge
    t\pr{A}-\frac{t^2}{2\mu}.
\end{align*}
Taking $t=\mu\pr{A}$ yields the first inequality.
We can then divide by$\pr{A}$ to obtain the second inequality.
\end{proof}

The next lemma relates gains from trade between
independent draws from a log-concave distribution
to that distribution's standard deviation. We
will use its upper bound for the original market
and its lower bound for the belief-robust market.

\begin{lemma}\label{lem:dispersion}
    Let $Z$ be an integrable random variable with
    a log-concave density, and let $Z'$ be an
    independent copy. Then $Z$ has finite variance,
    and
    \begin{equation}\label{eq:dispersion}
        \frac{\sqrt{\operatorname{Var}(Z)}}2
        \le
        \ex{(Z-Z')_+}
        \le
        \frac{\sqrt{\operatorname{Var}(Z)}}{\sqrt2}.
    \end{equation}
    The same inequalities hold when $Z$ is constant.
\end{lemma}

\begin{proof}
If $Z$ is constant, all three quantities are zero.
Otherwise, log-concavity is preserved by independent
products and affine transformations, so $Z-Z'$ has
a symmetric log-concave density. It follows that
$|Z-Z'|$ has a log-concave density on $[0,\infty)$,
and hence is MHR.

We now apply the moment comparison of Barlow,
Marshall, and Proschan
\cite[Theorem~4.3]{barlowmarshallproschan1963},
which we restate here:

\begin{theorem}[Barlow--Marshall--Proschan]
\label{thm:mhr-moment}
    Let $T$ be a nonnegative random variable with
    a density, a nondecreasing hazard rate, and
    finite mean. Then $T$ has finite second moment,
    and
    \[
        \ex{T^2}\le 2\bigl(\ex{T}\bigr)^2.
    \]
\end{theorem}

Applying Theorem~\ref{thm:mhr-moment} to $|Z-Z'|$,
which is nonnegative, integrable, and MHR, gives
\[
    \ex{(Z-Z')^2}
    \le
    2\bigl(\ex{|Z-Z'|}\bigr)^2
    <\infty.
\]

It follows that 
\[
    \operatorname{Var}(Z)
    =
    \ex{\bigl(\ex{Z-Z'\mid Z}\bigr)^2}
    \le
    \ex{(Z-Z')^2},
\]
by conditional Jensen's inequality. Thus, $Z$ has finite variance. As $Z$ and $Z'$ are independent,
\[
    \ex{(Z-Z')^2}=2\operatorname{Var}(Z).
\]

By symmetry,
$\ex{(Z-Z')_+}=\frac12\ex{|Z-Z'|}$.
Putting everything together, we find that
\[
    \ex{(Z-Z')_+}
    \ge
    \frac{\sqrt{\operatorname{Var}(Z)}}2.
\]
For the upper bound, we can apply Cauchy--Schwarz to find
\[
    \ex{(Z-Z')_+}
    =
    \frac12\ex{|Z-Z'|}
    \le
    \frac12\sqrt{\ex{(Z-Z')^2}}
    =
    \frac{\sqrt{\operatorname{Var}(Z)}}{\sqrt2}.
\]
\end{proof}

Putting these results together to prove the result, first, apply Lemma~\ref{lem:forecast} to the event
$\{\Sigma_S=\sigma_S^-\}$. By the definition of $d$
and the rarity assumption,
\begin{equation}\label{eq:pessimistic-forecast}
    d
    =
    \ex{C\mid\Sigma_S=\sigma_S^-}
    \ge
    \frac{\mu}{2}\pr{\Sigma_S=\sigma_S^-}
    \ge
    \frac{\pi^-\mu}{2}.
\end{equation}
It remains to translate this forecast bound into
a comparison between $R$ and $\mathrm{FB}$.

We begin by upper bounding the original first-best.
Because $C\ge0$, 
\begin{align}
    \mathrm{FB}
    &=
    \ex{
        \bigl(
            C(x(\theta_B)-x(\theta_S))
            +y(\theta_B)-y(\theta_S)
        \bigr)_+
    }
    \notag\\
    &\le
    \ex{C(x(\theta_B)-x(\theta_S))_+}
    +
    \ex{(y(\theta_B)-y(\theta_S))_+}
    \notag\\
    &=
    \mu\ex{(x(\theta_B)-x(\theta_S))_+}
    +
    \ex{(y(\theta_B)-y(\theta_S))_+}
    \notag\\
    &\le
    \frac{
        \mu\sqrt{\operatorname{Var}(x(\theta))}
        +\sqrt{\operatorname{Var}(y(\theta))}
    }{\sqrt2}.
    \label{eq:fb-dispersion-upper}
\end{align}
The final inequality
follows from Lemma~\ref{lem:dispersion}, since the
two marginal distributions are log-concave.

Next, we lower bound the variation in valuations
in the belief-robust market. Since $d\ge0$,
\begin{align*}
    \operatorname{Var}(\hat v_S)
    &=
    d^2\operatorname{Var}(x(\theta))
    +\operatorname{Var}(y(\theta))
    +2d\operatorname{Cov}(x(\theta),y(\theta))\\
    &\ge
    d^2\operatorname{Var}(x(\theta))
    +\operatorname{Var}(y(\theta))
    +2\rho d
    \sqrt{
        \operatorname{Var}(x(\theta))
        \operatorname{Var}(y(\theta))
    }\\
    &=
    \frac{1+\rho}{2}
    \left(
        d\sqrt{\operatorname{Var}(x(\theta))}
        +\sqrt{\operatorname{Var}(y(\theta))}
    \right)^2\\
    &\qquad+
    \frac{1-\rho}{2}
    \left(
        d\sqrt{\operatorname{Var}(x(\theta))}
        -\sqrt{\operatorname{Var}(y(\theta))}
    \right)^2\\
    &\ge
    \frac{1+\rho}{2}
    \left(
        d\sqrt{\operatorname{Var}(x(\theta))}
        +\sqrt{\operatorname{Var}(y(\theta))}
    \right)^2.
\end{align*}
The first inequality comes from the definition of $\rho$. The final inequality drops a nonnegative term,
since $\rho\le1$. This is where the correlation
restriction prevents the two components of value
from fully removing the idiosyncratic variation in valuations.

The belief-robust values are independent copies
of $dx(\theta)+y(\theta)$, which is log-concave
as an affine transformation of the preference
distribution. Applying the lower bound in
Lemma~\ref{lem:dispersion}, we find that
\begin{align}
    R
    &\ge
    \frac12\sqrt{\operatorname{Var}(\hat v_S)}
    \notag\\
    &\ge
    \frac{\sqrt{1+\rho}}{2\sqrt2}
    \left(
        d\sqrt{\operatorname{Var}(x(\theta))}
        +\sqrt{\operatorname{Var}(y(\theta))}
    \right).
    \label{eq:benchmark-dispersion}
\end{align}

We can now use \eqref{eq:pessimistic-forecast}.
Since $d\ge\pi^-\mu/2$ and $\pi^-\le1$,
\begin{align*}
    R
    &\ge
    \frac{\pi^-\sqrt{1+\rho}}{4\sqrt2}
    \left(
        \mu\sqrt{\operatorname{Var}(x(\theta))}
        +\sqrt{\operatorname{Var}(y(\theta))}
    \right)\\
    &\ge
    \frac{\pi^-}{4}\sqrt{1+\rho}\,
    \mathrm{FB},
\end{align*}
where the last inequality is
\eqref{eq:fb-dispersion-upper}.
This proves \eqref{eq:retention-target}.

Finally, applying Theorem~\ref{thm:reduction}, we find that
\[
    W^-_S(\mc I)
    \ge
    \frac{\pr{\Sigma_S=\sigma_S^-}}e
    \cdot\frac{R}{\mathrm{FB}}
    \ge
    \frac{(\pi^-)^2}{4e}\sqrt{1+\rho},
\]
as desired.
\end{proof}

%% file: twosided.tex
\section{Two-Sided Information}\label{sec:two-sided}

In this section, we generalize beyond the setting with an informed seller who proposes to an uninformed buyer to the case where both the seller and buyer may be differently informed. We keep the same mechanism, with the seller proposing to the buyer. This adds one additional core complexity; previously, the seller's quote may impact the buyer's valuation for the seller's asset. Now, in addition to this, the buyer's choice to accept or reject the quote affects the seller's valuation for the asset.

\paragraph{An informed buyer changes the seller's valuation upon acceptance.}To illustrate this complexity, consider an example with purely multiplicative
valuations $v(c,\theta)=cx(\theta)$, where $C$ takes values
$1$ and $3$ with equal probability, and $x(\theta)$ is uniformly
distributed on $[1,2]$. The seller is uninformed and proposes,
while the buyer observes $C$.

Consider a seller with $x(\theta_S)=1$, whose initial expected
value for the asset is $\ex{C}=2$. If the common component were
known to equal $2$, this seller would optimally quote $3$:
the buyer's value would be uniform on $[2,4]$, and each sale
would yield a profit of $1$. In the actual market, however,
the buyer's value lies in $[1,2]$ when $C=1$ and in $[3,6]$
when $C=3$. Consequently, a quote of $3$ is accepted only
when $C=3$. Conditional on acceptance, the seller's expected
value for the asset is therefore $3$, not $2$, and the sale
yields no profit.

The seller instead optimally quotes $\frac92$. At this price,
trade still occurs only when $C=3$, but the price now exceeds
the seller's value conditional on trade. Half of the good-news
buyers accept, so the seller's expected profit is
\[
    \frac12\cdot\frac12\cdot\left(\frac92-3\right)
    =\frac38.
\]\\

Because buyer acceptance may revise the seller's valuation for the asset, our original deviation argument needs to be amended. Previously, we constructed a reduction to an IPV market where both the buyer and seller publicly observe the most pessimistic seller signal. This is no longer acceptable as a lower bound, as the seller's profit from quoting in the original market may now fall below the profit they would realize upon making the same quote in the IPV market due to revisions that may occur in their posterior for the common component. To repair this, we have to consider a generalization of our IPV market from Section \ref{sec:reduction} that is robust to upwards revisions in the seller's posterior.

\subsection{Two-Sided Belief-Robust Values}

We obtain the correct generalization by considering the IPV market that arises when both parties publicly observe the joint signal pair $(\sigma_S^-, \sigma_B^+)$, where $\sigma_S^-$ is the most pessimistic seller signal (as we had before), and $\sigma_B^+$ is the most optimistic buyer signal. This repairs the argument, as the seller's valuation cannot be revised further upwards conditional on the realization of the most optimistic buyer signal; as such, the profit the seller obtains in this IPV market again serves as a valid lower bound on profit obtained in the original market. Downwards revisions do not affect the validity of the lower bound; a downward revision only increases the seller's profit conditional on buyer acceptance.

Formally, we define
\begin{equation}
    \sigma^-_S \in \argmin_{\sigma_S \in \mc S_S}
    m(\sigma_S, \sigma_B^*),
    \qquad
    \sigma^+_B \in \argmax_{\sigma_B \in \mc S_B}
    m(\sigma_S^*, \sigma_B)
\end{equation}
By our ordered news assumption, there exists a fixed $\sigma_S^-$ and $\sigma_B^+$ that satisfy this for all $(\sigma_S^*, \sigma_B^*) \in \mc S_S \times \mc S_B$ simultaneously. We then define the IPV market analogously to how it was defined in Section \ref{sec:reduction}, by defining $d := m(\sigma_S^-, \sigma_B^+)$, and considering valuations assuming a publicly observed forecast of $C$ given by $d$:
\begin{equation}
    \hat v_S := dx(\theta_S) + y(\theta_S),
    \qquad
    \hat v_B := dx(\theta_B) + y(\theta_B)
\end{equation}
As before, first-best gains from trade in this IPV market are then given by
\begin{equation}
    R := \ex{\pa{\hat v_B - \hat v_S}_+}
\end{equation}
Notice that if the buyer is uninformed, we exactly recover the reduction from Section \ref{sec:reduction}.


\begin{theorem}\label{thm:two-reduction}
    Let $\mc I := (v,\mc G,\mc F,\Sigma_S,\Sigma_B)$ be an
    instance. Suppose that $\ex{|C|}<\infty$,
    $\mathrm{FB}>0$, and, under $\mc G$, $x(\theta)$ and
    $y(\theta)$ have finite first moments and a jointly
    log-concave density. Then
    \begin{equation}
        W^-_S(\mc I)
        \ge
        \frac{\pr{\Sigma_S=\sigma_S^-, \Sigma_B=\sigma_B^+}}{e}
        \cdot\frac{R}{\mathrm{FB}}.
    \end{equation}
\end{theorem}

\begin{proof}
The fine-menu and screening arguments from
Section~\ref{sec:reduction} can be directly applied without modification.
We therefore only need to adapt the seller's
deviation argument. We will show that a seller
who observes $\sigma_S^-$ may secure at least
$\pr{\Sigma_B=\sigma_B^+\mid\Sigma_S=\sigma_S^-}$
times her monopoly profit in the belief-robust
market. Averaging over these sellers then gives
the joint probability appearing in the theorem.

Fix a finite price menu $\mc Q=Q\cup\{\bot\}$ and
an equilibrium
$(p,a,\lambda)\in\mc E_S(\mc I,\mc Q)$. We use
$u_{\mc Q}$ and $u$ as defined in
\eqref{eq:grid-profit} and
\eqref{eq:continuous-profit}, now with the
belief-robust valuations defined in this section.

Notice first that a buyer with signal
$\sigma_B^+$ uses, at every on- or off-path
quote, a mixture of the forecasts
$m(\sigma_S,\sigma_B^+)$. By the common news
ordering, each of these forecasts is at least
$m(\sigma_S^-,\sigma_B^+)=d$. Since
$x(\theta_B)\ge0$, the buyer's expected valuation
is therefore at least
$dx(\theta_B)+y(\theta_B)=\hat v_B$.
As such, any buyer with signal $\sigma_B^+$ and
$\hat v_B>q$ strictly prefers to accept the quote
$q$, regardless of the permissible beliefs
$\lambda$. Consequently,
\[
    \ex{a(\theta_B,\sigma_B^+,q)}
    \ge \pr{\hat v_B>q},
\]
where the expectation on the left is over
$\theta_B\sim\mc G$.

Now consider a seller with signal $\sigma_S^-$
and type $\theta$, and write
$k=dx(\theta)+y(\theta)$ for her value in the
belief-robust market. Suppose that this seller
quotes an available price $q>k$.
For every buyer signal $\sigma_B$, the buyer's expected value of
retaining the asset is
\begin{align*}
    &\ex{
        v(C,\theta)
        \mid
        \Sigma_S=\sigma_S^-,
        \Sigma_B=\sigma_B
    }\\
    &\qquad=
    m(\sigma_S^-,\sigma_B)x(\theta)+y(\theta)
    \le dx(\theta)+y(\theta)
    =k.
\end{align*}
Here the equality uses the affine value
function, and the inequality uses the fact
that $\sigma_B^+$ is the most optimistic buyer
report and $x(\theta)\ge0$.

Unlike in the one-sided argument, acceptance
may convey information about $C$ through the
buyer's signal. However, once both signals
are fixed, acceptance of this fixed quote
depends only on the buyer's type and
randomization. These are independent of $C$
conditional on the signals. It follows that
\begin{align*}
    &\ex{
        (q-v(C,\theta))a(\theta_B,\sigma_B,q)
        \mid
        \Sigma_S=\sigma_S^-,
        \Sigma_B=\sigma_B
    }\\
    &\qquad=
    \bigl(q-v(m(\sigma_S^-,\sigma_B),\theta)\bigr)
    \ex{a(\theta_B,\sigma_B,q)}
    \ge0.
\end{align*}
Thus, conditional on any buyer signal realization
that can occur alongside $\sigma_S^-$, the seller's
expected profit from this deviation is nonnegative.
We may therefore lower bound expected seller profit
by retaining only the contribution from buyers
whose signal is $\sigma_B^+$.

When $\Sigma_B = \sigma_B^+$, the seller's
expected value is exactly $k$, and the
acceptance probability is at least
$\pr{\hat v_B>q}$. Consequently,
\begin{align*}
    &\ex{
        (q-v(C,\theta))a(\theta_B,\Sigma_B,q)
        \mid \Sigma_S=\sigma_S^-
    }\\
    &\qquad\ge
    \pr{\Sigma_B=\sigma_B^+\mid\Sigma_S=\sigma_S^-}
    (q-k)\pr{\hat v_B>q}.
\end{align*}
Because the seller may also exit, equilibrium
best response guarantees expected utility of at least
\[
    \pr{\Sigma_B=\sigma_B^+\mid\Sigma_S=\sigma_S^-}
    \cdot u_{\mc Q}(k).
\]

Conditional on $\Sigma_S=\sigma_S^-$, the
seller's type still has distribution $\mc G$.
Averaging over these sellers and multiplying
by $\pr{\Sigma_S=\sigma_S^-}$ therefore gives
aggregate seller profit of at least
\[
    \pr{\Sigma_S=\sigma_S^-,\Sigma_B=\sigma_B^+}
    \cdot\ex{u_{\mc Q}(\hat v_S)}.
\]
All other sellers obtain nonnegative expected
utility because they may exit. The buyer's
ex-ante expected utility is also nonnegative
by best response and on-path Bayesian
updating. Gains from trade equal the sum of
the traders' utilities, whence
\begin{equation}\label{eq:two-finite-sellerprofit}
    W_S(p,a,\lambda)\,\mathrm{FB}
    \ge
    \pr{\Sigma_S=\sigma_S^-,\Sigma_B=\sigma_B^+}
    \cdot\ex{u_{\mc Q}(\hat v_S)}.
\end{equation}

We can now apply the remaining arguments from
Section~\ref{sec:reduction}. Take any fine menu
sequence $(\mc Q_n)$. Since
\eqref{eq:two-finite-sellerprofit} holds for
every equilibrium on each menu, taking the
infimum over equilibria, then the lower limit
over menus, and applying
Lemma~\ref{lem:finemenu} gives
\begin{align*}
    &\mathrm{FB}\cdot
    \liminf_{n\to\infty}
    W^-_S(\mc I,\mc Q_n)\\
    &\qquad\ge
    \pr{\Sigma_S=\sigma_S^-,\Sigma_B=\sigma_B^+}
    \cdot\ex{u(\hat v_S)}.
\end{align*}

Finally, $\hat v_B=dx(\theta_B)+y(\theta_B)$
is integrable and has a log-concave survival
function, by the same affine-projection
argument used in Section~\ref{sec:reduction}.
Lemma~\ref{lem:screening}, together with the
independence of $\hat v_S$ and $\hat v_B$,
therefore yields
\[
    \ex{u(\hat v_S)}
    \ge
    \frac1e\ex{(\hat v_B-\hat v_S)_+}
    =\frac{R}{e}.
\]
Combining the last two inequalities, dividing
by $\mathrm{FB}$, and taking the infimum over
fine menu sequences proves the theorem.
\end{proof}

\subsection{Lower Bound on Equilibrium Gains From Trade}

To obtain a lower bound on equilibrium gains from trade, we proceed in a very similar manner as in Section \ref{sec:retention}. The main difference is that we now need to control two parameters: the rarity of the most pessimistic seller signal $\sigma_S^-$ as we had before, and the rarity of the most optimistic buyer signal $\sigma_B^+$. Under identical regularity as in Section \ref{sec:retention}, this implies a lower bound on equilibrium gains from trade.

We define this regularity as follows.

\begin{definition}[$(\pi^-, \pi^+, \rho)$-admissible instance]
    An instance $(v, \mc G, \mc F, \Sigma)$ is
    $(\pi_-, \pi^+, \rho)$-admissible if
    \begin{enumerate}
        \item $\mc F$ has non-negative support and is
        dominated in convex order by an exponential
        variable of mean
        $\mu := \exx{\mc F}{C} \in (0, \infty)$.
        That is, for any convex function
        $\phi: \R \to \R$,
        \[
            \ex{\phi(C)} \le \ex{\phi(Z)}
        \]
        where $Z$ is an exponential variable with mean $\mu$.

        \item The rarity of the most pessimistic seller
        signal is lower bounded uniformly over
        realizations of the buyer's signal:
        \[
            \min_{\sigma_B \in \mc S_B}
            \pr{
                \Sigma_S = \sigma_S^-
                \mid \Sigma_B = \sigma_B
            }
            \ge \pi^- > 0
        \]

        \item The rarity of the most optimistic buyer
        signal is lower bounded uniformly over
        realizations of the seller's signal:
        \[
            \min_{\sigma_S \in \mc S_S}
            \pr{
                \Sigma_B = \sigma_B^+
                \mid \Sigma_S = \sigma_S
            }
            \ge \pi^+ > 0
        \]

        \item For $\theta \sim \mc G$,
        $(x(\theta), y(\theta))$ has a log-concave
        density and
        $\ex{x(\theta) + \abs{y(\theta)}} < \infty$

        \item $\operatorname{Corr}(x(\theta), y(\theta))
        \ge \rho > -1$.
        If either $x(\theta)$ or $y(\theta)$ is constant,
        then take this to be satisfied for $\rho = 0$.
    \end{enumerate}
\end{definition}

We are now ready to state the lower bound.
\begin{theorem}\label{thm:two-retention}
For any $(\pi^-, \pi^+, \rho)$-admissible instance
$\mc I$ in which the seller proposes,
\[
    W_S^-(\mc I)
    \ge
    \frac{(\pi^-)^2 \pi^+}{4e}
    \sqrt{1 + \rho}
\]
\end{theorem}

\begin{proof}
The reduction and screening arguments have already
been established by Theorem~\ref{thm:two-reduction}.
It therefore remains to lower bound
$R/\mathrm{FB}$ and the probability of the
selected signal pair.

Notice first that averaging the seller's
conditional rarity bound over the buyer's signal
gives
\begin{align*}
    \pr{\Sigma_S=\sigma_S^-}
    &=
    \sum_{\sigma_B\in\mc S_B}
    \pr{\Sigma_B=\sigma_B}
    \pr{\Sigma_S=\sigma_S^-\mid\Sigma_B=\sigma_B}\\
    &\ge \pi^-.
\end{align*}
Together with the buyer's conditional rarity
bound, this implies
\begin{align}
    &\pr{\Sigma_S=\sigma_S^-,\Sigma_B=\sigma_B^+}
    \notag\\
    &\qquad=
    \pr{\Sigma_S=\sigma_S^-}
    \pr{\Sigma_B=\sigma_B^+\mid\Sigma_S=\sigma_S^-}
    \ge \pi^-\pi^+.
    \label{eq:two-joint-probability}
\end{align}

We next lower bound $d$. By the common news
ordering, $m(\sigma_S^-,\sigma_B)\le d$ for every
buyer signal realization. Averaging over these realizations
therefore gives
\[
    d\ge \ex{C\mid\Sigma_S=\sigma_S^-}.
\]
We may now apply Lemma~\ref{lem:forecast} to the
event $\{\Sigma_S=\sigma_S^-\}$ to obtain
\begin{equation}\label{eq:two-forecast}
    d
    \ge
    \ex{C\mid\Sigma_S=\sigma_S^-}
    \ge
    \frac{\mu}{2}\pr{\Sigma_S=\sigma_S^-}
    \ge
    \frac{\mu\pi^-}{2}.
\end{equation}
Notice that we apply the lemma to the seller's
signal alone, rather than to the joint signal
pair. The optimistic buyer signal can only
increase the forecast relative to this
conditional average.

We now reuse the dispersion comparison from
Section~\ref{sec:retention}. Its derivation from
Lemma~\ref{lem:dispersion} and the correlation
assumption applies to any nonnegative forecast
$d$. Consequently,
\begin{align}
    \frac{R}{\mathrm{FB}}
    &\ge
    \frac{\sqrt{1+\rho}}{2}
    \cdot
    \frac{
        d\sqrt{\operatorname{Var}(x(\theta))}
        +\sqrt{\operatorname{Var}(y(\theta))}
    }{
        \mu\sqrt{\operatorname{Var}(x(\theta))}
        +\sqrt{\operatorname{Var}(y(\theta))}
    }
    \notag\\
    &\ge
    \frac{\pi^-}{4}\sqrt{1+\rho}.
    \label{eq:two-retention}
\end{align}
The last inequality follows from
\eqref{eq:two-forecast} and $\pi^-\le1$:
the numerator is at least $\pi^-/2$ times
the denominator.

Finally, applying
Theorem~\ref{thm:two-reduction} and combining
\eqref{eq:two-joint-probability} with
\eqref{eq:two-retention}, we find that
\begin{align*}
    W^-_S(\mc I)
    &\ge
    \frac{
        \pr{\Sigma_S=\sigma_S^-,\Sigma_B=\sigma_B^+}
    }{e}
    \cdot\frac{R}{\mathrm{FB}}\\
    &\ge
    \frac{\pi^-\pi^+}{e}
    \cdot\frac{\pi^-}{4}\sqrt{1+\rho}\\
    &=
    \frac{(\pi^-)^2\pi^+}{4e}\sqrt{1+\rho},
\end{align*}
as desired.
\end{proof}

%% file: mechanisms.tex
\section{Alternate Mechanisms}\label{s:mechanisms}

Thus far, we have analyzed the seller-proposing mechanism. In this section, we give a similar analysis for two additional mechanisms: the buyer proposing mechanism $P = B$, and the fair random-offerer mechanism, where $P$ is equiprobably selected to be either $B$ or $S$. For the buyer-proposing mechanism, we obtain the exact same bound as in Theorem \ref{thm:two-reduction}. For the random-offerer mechanism, we obtain the exact bound as Theorem \ref{thm:two-reduction} as well, but under substantial weaker regularity conditions over the joint distribution of $(x(\theta), y(\theta))$ than joint log-concavity. 

\subsection{The Buyer-Proposing Mechanism}

Our final lower bound of Theorem \ref{thm:two-retention} for the seller-proposing mechanism was handled in two steps: the deviation argument of Theorem \ref{thm:two-reduction} and then an argument showing that the first-best gains from trade in our IPV market reduction is not vanishingly small relative to the original first-best. Only the deviation argument depended on $P$. We prove the same final lower bound for the buyer-proposing mechanism by showing an analogous deviation argument for a proposing buyer. 

\begin{theorem}\label{thm:two-retention-buyer}
For any $(\pi^-, \pi^+, \rho)$-admissible instance
$\mc I$ in which the buyer proposes,
\[
    W_B^-(\mc I)
    \ge
    \frac{(\pi^-)^2 \pi^+}{4e}
    \sqrt{1 + \rho}
\]
\end{theorem}
\begin{proof}
We use the same belief-robust market as in
Theorem~\ref{thm:two-reduction}, with
$d=m(\sigma_S^-,\sigma_B^+)$ and valuations
$\hat v_S=dx(\theta_S)+y(\theta_S)$ and
$\hat v_B=dx(\theta_B)+y(\theta_B)$.
We first show that a buyer who observes $\sigma_B^+$
can guarantee a fraction of her optimal profit in this
market. We then apply our existing screening and
retention bounds.

For a finite price menu $\mc Q=Q\cup\{\bot\}$ and a
buyer with value $k$, define
\[
    u^B_{\mc Q}(k)
    :=
    \max\left(
        \{0\}\cup
        \left\{
            (k-q)\pr{\hat v_S<q}
            :q\in Q,\ q<k
        \right\}
    \right).
\]
This is the buyer's optimal profit in the
belief-robust market, where we again require strict
acceptance so that the guarantee does not depend
on the receiver's tie-breaking.

Fix an equilibrium $(p,a,\lambda)$.
Notice first that, at every on- or off-path bid,
a seller who observes $\sigma_S^-$ uses a mixture
of forecasts
\[
    m(\sigma_S^-,\sigma_B)\le d.
\]
Since $x(\theta_S)\ge0$, her expected valuation
is therefore at most
$dx(\theta_S)+y(\theta_S)=\hat v_S$.
As such, she strictly prefers to accept any bid
$q>\hat v_S$, regardless of her beliefs about
the buyer's signal.

Now consider a buyer with signal $\sigma_B^+$ and
type $\theta$, and write $k=dx(\theta)+y(\theta)$.
By the common news ordering,
\[
    m(\sigma_S,\sigma_B^+)\ge d
\]
for every seller signal compatible with
$\sigma_B^+$.
Consequently, at any bid $q<k$, this buyer's
expected valuation conditional on either party's
signal is at least $k>q$.
In particular, acceptance by sellers with other
signals cannot contribute negative expected
profit to this deviation. To see this explicitly,
independence of preferences from the common
component and signals gives
\begin{align*}
    &\ex{
        \bigl(v(C,\theta)-q\bigr)
        a(\theta_S,\sigma_S,q)
        \,\middle|\,
        \Sigma_S=\sigma_S,\Sigma_B=\sigma_B^+
    }\\
    &\qquad=
    \bigl(
        m(\sigma_S,\sigma_B^+)x(\theta)+y(\theta)-q
    \bigr)
    \ex{a(\theta_S,\sigma_S,q)}
    \ge0.
\end{align*}
When the seller's signal is $\sigma_S^-$,
the factor in parentheses is exactly $k-q$,
and the seller accepts whenever $\hat v_S<q$.
Keeping just the contribution from this signal,
the buyer therefore obtains expected profit
at least
\[
    \pr{\Sigma_S=\sigma_S^-\mid\Sigma_B=\sigma_B^+}
    (k-q)\pr{\hat v_S<q}.
\]
The buyer can also guarantee zero by exiting.
Best response thus guarantees expected utility
at least
\[
    \pr{\Sigma_S=\sigma_S^-\mid\Sigma_B=\sigma_B^+}
    u^B_{\mc Q}(k).
\]

We may now average over buyer types and signals.
Preferences are independent of the signals, and
buyers with every other signal obtain nonnegative
expected utility by their option to exit.
The seller's ex-ante expected utility is also
nonnegative, by receiver best response and
Bayes' rule on-path. Adding the parties'
utilities cancels the cash transfer, yielding
\begin{equation}\label{eq:buyer-grid-transfer}
    W_B(p,a,\lambda)
    \ge
    \frac{
        \pr{\Sigma_S=\sigma_S^-,\Sigma_B=\sigma_B^+}
    }{\mathrm{FB}}
    \ex{u^B_{\mc Q}(\hat v_B)}.
\end{equation}

It remains to pass to fine menus and apply the
screening bound. Define the continuous-price
analogue
\[
    u^B(k)
    :=
    \sup_{q<k}(k-q)\pr{\hat v_S<q}.
\]
Both steps follow from our existing lemmas by
reversing signs in the auxiliary pricing problem.
Specifically, a bid $q<k$ becomes an ask $-q>-k$
against receiver value $-\hat v_S$, with profit
\[
    (k-q)\pr{\hat v_S<q}
    =
    \bigl((-q)-(-k)\bigr)
    \pr{-\hat v_S>-q}.
\]
Negating every cash price preserves the fine-menu
property. Lemma~\ref{lem:finemenu}, applied to
receiver value $-\hat v_S$ and reservation value
$-\hat v_B$, therefore gives
\[
    \lim_{n\to\infty}
    \ex{u^B_{\mc Q_n}(\hat v_B)}
    =
    \ex{u^B(\hat v_B)}
\]
for every fine menu sequence $(\mc Q_n)$.
Since \eqref{eq:buyer-grid-transfer} holds for
every equilibrium on every such menu, it follows
that
\[
    W_B^-(\mc I)
    \ge
    \frac{
        \pr{\Sigma_S=\sigma_S^-,\Sigma_B=\sigma_B^+}
    }{\mathrm{FB}}
    \ex{u^B(\hat v_B)}.
\]

Next, $-\hat v_S=-dx(\theta_S)-y(\theta_S)$
is an affine projection of a log-concave
distribution, and therefore has a log-concave
survival function. Applying
Lemma~\ref{lem:screening} to this receiver value
and reservation value $-k$ gives
\[
    u^B(k)
    \ge
    \frac1e\ex{(k-\hat v_S)_+}.
\]
Because $\hat v_B$ and $\hat v_S$ are independent,
averaging over $k=\hat v_B$ yields
\[
    \ex{u^B(\hat v_B)}
    \ge
    \frac1e\ex{(\hat v_B-\hat v_S)_+}
    =
    \frac{R}{e}.
\]
We have therefore established the buyer-proposing
analogue of Theorem~\ref{thm:two-reduction}:
\begin{equation}\label{eq:buyer-reduction}
    W_B^-(\mc I)
    \ge
    \frac{
        \pr{\Sigma_S=\sigma_S^-,\Sigma_B=\sigma_B^+}
    }{e}
    \frac{R}{\mathrm{FB}}.
\end{equation}

Finally, the bounds
\eqref{eq:two-joint-probability} and
\eqref{eq:two-retention} depend only on the
instance and the belief-robust market, not on
which party proposes. They give
\[
    \pr{\Sigma_S=\sigma_S^-,\Sigma_B=\sigma_B^+}
    \ge\pi^-\pi^+,
    \qquad
    \frac{R}{\mathrm{FB}}
    \ge\frac{\pi^-}{4}\sqrt{1+\rho}.
\]
Substituting these into
\eqref{eq:buyer-reduction}, we obtain
\[
    W_B^-(\mc I)
    \ge
    \frac{(\pi^-)^2\pi^+}{4e}\sqrt{1+\rho},
\]
whence the desired result follows.
\end{proof}

\subsection{The Random-Offerer Mechanism}
We now consider a proposer $P^*$ that is drawn equiprobably from the set $\set{B, S}$. By drawing a random offerer, we are able to show a similar absolute constant bound as Theorem \ref{thm:two-retention} under weaker regularity. We are able to do so due to recent work that shows a distribution-free $\frac1e$ screening bound for IPV markets under a random offerer, due to Bei, Wu, and Zhou \cite[Theorem~3.5]{BWZ26}. We restate this result here, adapted for our use case:

\begin{theorem}[Profit comparison]\label{thm:profit-comparison}
Let $\hat v_B$ and $\hat v_S$ be independent,
nonnegative valuations with finite first moments.
Then
\[
    \frac12\ex{u(\hat v_S)}
    +
    \frac12\ex{u^B(\hat v_B)}
    \ge
    \frac1e\ex{(\hat v_B-\hat v_S)_+}.
\]
\end{theorem}

This allows us to prove the same bound but under weaker regularity conditions. Specifically, we can relax joint log-concavity of $(x(\theta), y(\theta))$ to marginal log-concavity. Further, if $x(\theta)$ and $y(\theta)$ are independent, the result holds even without any regularity assumptions on the distributions except integrability. We define this formally below:

\begin{definition}[$(\pi^-, \pi^+, \rho)$-weak admissible instance]
    An instance $(v, \mc G, \mc F, \Sigma)$ is
    $(\pi_-, \pi^+, \rho)$-weak admissible if
    \begin{enumerate}
        \item $\mc F$ has non-negative support and is
        dominated in convex order by an exponential
        variable of mean
        $\mu := \exx{\mc F}{C} \in (0, \infty)$.
        That is, for any convex function
        $\phi: \R \to \R$,
        \[
            \ex{\phi(C)} \le \ex{\phi(Z)}
        \]
        where $Z$ is an exponential variable with mean $\mu$.

        \item The rarity of the most pessimistic seller
        signal is lower bounded uniformly over
        realizations of the buyer's signal:
        \[
            \min_{\sigma_B \in \mc S_B}
            \pr{
                \Sigma_S = \sigma_S^-
                \mid \Sigma_B = \sigma_B
            }
            \ge \pi^- > 0
        \]

        \item The rarity of the most optimistic buyer
        signal is lower bounded uniformly over
        realizations of the seller's signal:
        \[
            \min_{\sigma_S \in \mc S_S}
            \pr{
                \Sigma_B = \sigma_B^+
                \mid \Sigma_S = \sigma_S
            }
            \ge \pi^+ > 0
        \]

     \item $\ex{x(\theta)+|y(\theta)|}<\infty$.

\item Either $x(\theta)$ and $y(\theta)$ are
independent, in which case we take $\rho=0$;
or their marginal distributions are log-concave
and
\[
    \operatorname{Corr}(x(\theta),y(\theta))
    \ge \rho>-1.
\]
    \end{enumerate}
\end{definition}

We are now ready to state the lower bound.
\begin{theorem}\label{thm:two-retention-random}
For any $(\pi^-,\pi^+,\rho)$-weak admissible instance
$\mc I$ with $\mathrm{FB}>0$, suppose that the
proposer $P^*$ is selected by an independent,
public fair coin. Then
\[
    W_{P^*}^-(\mc I)
    \ge
    \frac{(\pi^-)^2\pi^+}{8\sqrt3\,e}
    (1+\rho)^{3/2}.
\]
If $x(\theta)$ and $y(\theta)$ are independent,
the stronger bound
\[
    W_{P^*}^-(\mc I)
    \ge
    \frac{(\pi^-)^2\pi^+}{e(2+\pi^-)}
\]
holds without any shape restrictions on their
distributions.
\end{theorem}

\begin{proof}
We use the same belief-robust market as before,
with $d=m(\sigma_S^-,\sigma_B^+)$ and valuations
$\hat v_S=dx(\theta_S)+y(\theta_S)$ and
$\hat v_B=dx(\theta_B)+y(\theta_B)$.
We first show that random proposal guarantees
a $1/e$ fraction of this market's first-best,
weighted by the probability of the seller pessimistic/buyer optimistic joint
signal pair. We then compare this first-best
to that of the original market.

Notice that the deviation arguments establishing
\eqref{eq:two-finite-sellerprofit} and
\eqref{eq:buyer-grid-transfer} do not use
log-concavity. That assumption was only used
afterwards to bound the proposer profits
obtained through those deviations.
Averaging the two finite-menu guarantees over
the independent proposer draw and applying
Lemma~\ref{lem:finemenu}, we therefore obtain
\begin{align}
    W_{P^*}^-(\mc I)
    &\ge
    \frac{
        \pr{\Sigma_S=\sigma_S^-,
             \Sigma_B=\sigma_B^+}
    }{\mathrm{FB}}
    \left(
        \frac12\ex{u(\hat v_S)}
        +
        \frac12\ex{u^B(\hat v_B)}
    \right)
    \notag\\
    &\ge
    \frac{
        \pr{\Sigma_S=\sigma_S^-,
             \Sigma_B=\sigma_B^+}
    }{e}
    \frac{R}{\mathrm{FB}}.
    \label{eq:random-reduction}
\end{align}
The second inequality is
Theorem~\ref{thm:profit-comparison}.
The finite-menu guarantees hold uniformly over
the continuation equilibria under either
proposer, so this argument also applies to
worst equilibrium welfare along every fine
menu sequence.

The probability and forecast bounds from the
previous section likewise remain valid:
\begin{equation}\label{eq:random-corner-bounds}
    \pr{\Sigma_S=\sigma_S^-,
         \Sigma_B=\sigma_B^+}
    \ge \pi^-\pi^+,
    \qquad
    d\ge\frac{\pi^-\mu}{2}.
\end{equation}
Indeed, \eqref{eq:two-joint-probability} and
\eqref{eq:two-forecast} used only the news
ordering, rarity assumptions, and convex-order
domination of the common component.
It remains to lower bound $R/\mathrm{FB}$.

Suppose first that $x(\theta)$ and $y(\theta)$
are independent. Then
$x(\theta_B)-x(\theta_S)$ and
$y(\theta_B)-y(\theta_S)$ are independent,
integrable random variables with mean zero.
Conditional Jensen's inequality gives
\[
    R
    \ge
    d\,\ex{(x(\theta_B)-x(\theta_S))_+},
    \qquad
    R
    \ge
    \ex{(y(\theta_B)-y(\theta_S))_+}.
\]
For example, the first inequality follows by
conditioning on the difference in $x$ and
averaging over the mean-zero difference in $y$.
The second follows by reversing their roles.

We can then use those inequalties to bound first-best like so:
\begin{align*}
    \mathrm{FB}
    &\le
    \mu\ex{(x(\theta_B)-x(\theta_S))_+}
    +
    \ex{(y(\theta_B)-y(\theta_S))_+}\\
    &\le
    \left(\frac{\mu}{d}+1\right)R.
\end{align*}
Consequently,
\[
    \frac{R}{\mathrm{FB}}
    \ge
    \frac{d}{\mu+d}
    \ge
    \frac{\pi^-}{2+\pi^-}.
\]
Combining this with
\eqref{eq:random-reduction} and
\eqref{eq:random-corner-bounds} proves the
independent-components bound. Since $\rho=0$
in this case, it also implies the first bound
in the theorem.

Now suppose that the components are dependent
but have log-concave marginal distributions.
Their variances are finite by
Lemma~\ref{lem:dispersion}. For this part of the
proof, write
\[
    D:=\hat v_B-\hat v_S,
    \qquad
    s:=
    d\sqrt{\operatorname{Var}(x(\theta))}
    +
    \sqrt{\operatorname{Var}(y(\theta))}.
\]
Thus $R=\frac12\ex{|D|}$, since $D$ is the
difference of independent, identically
distributed valuations.

The variance calculation in the proof of
Theorem~\ref{thm:retention} used the correlation
restriction, but not joint log-concavity.
It therefore still gives
\begin{equation}\label{eq:random-second-moment}
    \ex{D^2}
    =
    2\operatorname{Var}(dx(\theta)+y(\theta))
    \ge
    (1+\rho)s^2.
\end{equation}
Unlike in that proof, we cannot apply
Lemma~\ref{lem:dispersion} directly to
$dx(\theta)+y(\theta)$: marginal log-concavity
does not imply log-concavity of this sum.
Instead, we control its fourth moment using
the marginal distributions.

For any nonconstant log-concave random variable
$Z$ and an independent copy $Z'$, the argument
in Lemma~\ref{lem:dispersion} shows that
$|Z-Z'|$ is MHR. The same moment comparison of
Barlow, Marshall, and Proschan
\cite[Theorem~4.3]{barlowmarshallproschan1963}
gives
\[
    \ex{(Z-Z')^4}
    \le
    6\bigl(\ex{(Z-Z')^2}\bigr)^2
    =
    24\operatorname{Var}(Z)^2.
\]
The inequality also holds when $Z$ is constant.
Applying it separately to $x(\theta)$ and
$y(\theta)$, followed by Minkowski's inequality,
yields
\begin{align*}
    \bigl(\ex{D^4}\bigr)^{1/4}
    &\le
    d\bigl(
        \ex{(x(\theta_B)-x(\theta_S))^4}
    \bigr)^{1/4}\\
    &\qquad+
    \bigl(
        \ex{(y(\theta_B)-y(\theta_S))^4}
    \bigr)^{1/4}\\
    &\le
    24^{1/4}s.
\end{align*}

We can now obtain a lower bound on $R$.
Applying H\"older's inequality, we get
\[
    \ex{D^2}
    \le
    \bigl(\ex{|D|}\bigr)^{2/3}
    \bigl(\ex{D^4}\bigr)^{1/3}.
\]
Rearranging and applying the preceding moment
bounds, we find that
\begin{align*}
    R
    &=
    \frac12\ex{|D|}\\
    &\ge
    \frac{
        \bigl(\ex{D^2}\bigr)^{3/2}
    }{
        2\bigl(\ex{D^4}\bigr)^{1/2}
    }\\
    &\ge
    \frac{(1+\rho)^{3/2}}{4\sqrt6}\,s.
\end{align*}

Finally, the upper bound
\eqref{eq:fb-dispersion-upper} only required
marginal log-concavity, so it remains valid:
\[
    \mathrm{FB}
    \le
    \frac{
        \mu\sqrt{\operatorname{Var}(x(\theta))}
        +
        \sqrt{\operatorname{Var}(y(\theta))}
    }{\sqrt2}.
\]
Consequently,
\begin{align*}
    \frac{R}{\mathrm{FB}}
    &\ge
    \frac{(1+\rho)^{3/2}}{4\sqrt3}
    \frac{
        d\sqrt{\operatorname{Var}(x(\theta))}
        +
        \sqrt{\operatorname{Var}(y(\theta))}
    }{
        \mu\sqrt{\operatorname{Var}(x(\theta))}
        +
        \sqrt{\operatorname{Var}(y(\theta))}
    }\\
    &\ge
    \frac{\pi^-}{8\sqrt3}(1+\rho)^{3/2}.
\end{align*}
The last inequality uses
$d\ge\pi^-\mu/2$ and $\pi^-\le1$.
Combining this with
\eqref{eq:random-reduction} and
\eqref{eq:random-corner-bounds} gives
\[
    W_{P^*}^-(\mc I)
    \ge
    \frac{(\pi^-)^2\pi^+}{8\sqrt3\,e}
    (1+\rho)^{3/2},
\]
whence the desired result follows.
\end{proof}


%% file: correlation_example.tex
\section{Necessity of Restricted Correlation}\label{app:correlation}

The following example shows that a correlation restriction on the two
components~$x,y$ of an agent's private type is necessary for good
worst-case equilibrium approximation bounds.

\begin{example}[Negative Correlation Can Limit Equilibrium Gains from Trade]
\label{ex:correlation}
\mbox{}\\

\vspace{-.5\baselineskip}
\noindent
\textbf{The setup:}
We now consider affine valuations $cx+y$ rather than purely
multiplicative valuations.
The example is parameterized by $\epsilon \in (0,\tfrac12]$.
Each trader's private type is a pair $(x,y)$: first draw $x$ uniformly
from $[0,1]$, and then draw $y$ uniformly from
$[1-x,1-x+\epsilon]$.
The buyer's and seller's types are drawn independently.
The common component~$c$ is drawn uniformly from $[0,2]$,
independently of both traders' types.

The joint distribution of $(x,y)$ is uniform on a convex
parallelogram and hence has a log-concave density.
All valuations are nonnegative.
However, $x$ and $y$ are negatively correlated, with
\[
    \rho := \operatorname{Corr}(x,y)
    = -\frac{1}{\sqrt{1+\epsilon^2}}.
\]
Thus, as $\epsilon$ approaches zero, the correlation approaches~$-1$:
a larger multiplicative component~$x$ is almost exactly offset by a
smaller additive component~$y$.

\vspace{.5\baselineskip}
\noindent
\textbf{The first-best benchmark:}
The expected maximum gains from trade are at least $1/12$,
independently of~$\epsilon$.
Indeed, conditional on $c$ and $x$, a trader's expected valuation is
\[
    cx+\expect{y\mid x}
    =
    1+(c-1)x+\frac{\epsilon}{2}.
\]
Thus, allocating the item to the trader with larger~$x$ when $c>1$,
and to the trader with smaller~$x$ when $c<1$, yields expected gains
from trade
\[
    \expect{\bigl((c-1)(x_B-x_S)\bigr)_+}
    =
    \expect{|c-1|}\cdot\expect{(x_B-x_S)_+}
    =
    \frac12\cdot\frac16
    =
    \frac1{12}.
\]
The first-best allocation can additionally use the realized additive
components $y_S$ and $y_B$, so its expected gains are at least as
large.
As $\epsilon$ approaches zero, valuations converge to
$1+(c-1)x$, and the first-best benchmark converges to $1/12$.

\vspace{.5\baselineskip}
\noindent
\textbf{The information structure:}
Neither trader observes any information about~$c$.
Equivalently, each receives a single signal that occurs with
probability~$1$.
Thus, there are no rare signals in this example.
Both traders know that $\expect{c}=1$, so a trader with type $(x,y)$
has expected value $x+y$ for the item.
These expected values are independently and uniformly distributed
on $[1,1+\epsilon]$.

\vspace{.5\baselineskip}
\noindent
\textbf{An equilibrium:}
Because the seller has no information about~$c$, its price cannot
communicate any such information to the buyer.
The buyer therefore accepts an offered price~$p$ if and only if
$x_B+y_B \ge p$.
The seller sets a monopoly price, maximizing~\eqref{eq:intro} with
$v_S$ identified with its expected value $x_S+y_S$, and with the
buyer value distribution~$F_B$ identified with the uniform
distribution on $[1,1+\epsilon]$.
That is:
\begin{equation}\label{eq:correlation}
    p(x_S,y_S)
    :=
    \frac{1+\epsilon+x_S+y_S}{2}.
\end{equation}
This price maximizes the seller's expected utility given the
buyer's acceptance rule.
It is also at least the seller's expected value $x_S+y_S$, so
the seller cannot benefit from refusing to sell.
The buyer's forecast of~$c$ remains~$1$ at every price, including
unused prices.

\vspace{.5\baselineskip}
\noindent
\textbf{Equilibrium gains from trade:}
The expected gains from trade under~\eqref{eq:correlation} are
$\epsilon/8$.
Thus, seller proposing obtains a constant $3/4$ fraction of the
IPV market's first-best gains from trade, but a vanishing fraction
of the original first-best.
Since the latter benchmark converges to $1/12$,
\[
    \frac{\text{equilibrium gains from trade}}
         {\text{original first-best gains from trade}}
    \sim
    \frac{\epsilon/8}{1/12}
    =
    \frac32\epsilon
    \qquad\text{as $\epsilon\to0$.}
\]
The loss here is not due to screening: screening retains a constant
fraction of the gains available given the traders' information.
Rather, that information permits only a vanishing fraction of the
original first-best gains from trade.\footnote{Indeed, any
allocation based only on the traders' available information obtains
expected gains from trade at most $\epsilon/6$.
Thus, even the best equilibrium obtains a vanishing fraction of the
original first-best benchmark.}

To understand why, consider the limiting case $y=1-x$.
A trader's valuation is then $1+(c-1)x$.
When $c>1$, the trader with larger~$x$ values the item more;
when $c<1$, the trader with smaller~$x$ values it more.
The first-best allocation can respond to this reversal.
The traders cannot: without observing~$c$, their expected valuations
are all equal to~$1$.
For small positive~$\epsilon$, these expected valuations differ
only slightly, even though substantial gains remain available to
an allocation that observes~$c$.

Finally, the rate of decay matches the correlation dependence in
our lower bounds.
Since
\[
    1+\rho
    =
    1-\frac{1}{\sqrt{1+\epsilon^2}}
    \sim \frac{\epsilon^2}{2},
\]
we have
\[
    \frac{\text{equilibrium gains from trade}}
         {\text{original first-best gains from trade}}
    \sim
    \frac{3}{\sqrt2}\sqrt{1+\rho}.
\]
Thus, the $\sqrt{1+\rho}$ dependence in Theorem \ref{thm:two-retention} is tight up to an absolute
multiplicative constant, even with no rare signals and a fixed
distribution of the common component.

\end{example}

%% file: appendix.tex
\section{Proof of Theorem \ref{thm:eqexist}}

The high-level proof approach to show existence of an equilibrium $(p, a, \lambda)$ for any market instance $\mc I$ and finite price menu $\mc Q$ is to first approximate our continuous type space by a finite type space. We can then apply the equilibrium existence result of Kreps and Wilson in their original finite-type formulation \cite[Proposition 1]{krepswilson1982}. To conclude, it suffices to show that finer approximations yield an equilibrium under continuous types.

Formally, we begin by choosing finite-valued measurable functions $x_n, y_n$ that converge pointwise to $x$ and $y$ respectively (e.g. by truncating and rounding). We then construct a finite type game as follows. Rather than the original type space $\Theta$, let $\Theta'_n := \set{(x_n(\theta), y_n(\theta)}$ denote our new finite type space. In this finite version of the game, we also let types be drawn i.i.d according to the pushforward of $\mc G$ under $(x_n, y_n)$. Finally, let valuations be given by $m(\sigma_S, \sigma_B) x_n(\theta) + y_n(\theta)$. Because we have finite signal spaces $\mc S_S$ and $\mc S_B$, the two-stage proposer game is now a fully finite extensive-form game with perfect recall. We can consequently invoke \cite[Proposition 1]{krepswilson1982} to find a sequential equilibrium $(p^n, a^n, \lambda^n)$ for this finite approximation.

We can now lift these strategies to $\Theta$ by assigning each original type to the strategy of its approximating type $(x^n(\theta), y^n(\theta)) \in \Theta'_n$. As $(p^n, a^n, \lambda^n)$ is a sequential equilibrium in the finite game, there exist mixed proposer strategies with full support that converge to $p^n$. Averaging those over the proposer's independent type $\theta_P$, we obtain a sequence of strictly positive quote probabilities $\ell_q^{n, k}$ that sum to one over $q \in Q$, and receiver posteriors 
\begin{equation}
\pr{\Sigma_P = \sigma_P \mid \Sigma_R = \sigma_R, p^{n, k}(\theta_P, \sigma_P) = q} = \frac{\ell^{n, k}_q(\sigma_P) \cdot \pr{\Sigma_P = \sigma_P \mid \Sigma_R = \sigma_R}}{\sum_{\sigma_P'}\ell^{n, k}_q(\sigma_P') \cdot \pr{\Sigma_P = \sigma_P' \mid \Sigma_R = \sigma_R}}
\end{equation}
following from Bayes' rule. The limits of these posteriors yield $\lambda^n$ and therefore satisfy conditions 3 and 4 of our sequential equilibrium concept. Next, define
\[\ell_q^n(\sigma_P) := \int p_q^n(\theta, \sigma_P d\mc G(\theta), \qquad \alpha_{\sigma_R, q}^n := \int a^n(\theta, \sigma_R, q) d \mc G(\theta)\]
Because these, along with $\lambda^n$, are each finite vectors, we may pass to a subsequence along with they converge they converge to some $\ell ,\alpha, \lambda$. Our goal now is to recover proposer and receiver best responses from these limiting objects.

To do this, we use the following lemma
\begin{lemma}
	Let $\mc D$ be a finite action set with measurable, real-valued payoffs $f_d^n(\theta) \to f_d(\theta)$ for every $d$, almost everywhere. Suppose measurable mixed strategies $b^n$ use only maximizers of $f^n$, almost everywhere, and $\int b^n_d d\mc G \to \beta_d$. Then, there exists a measurable mixed strategy $b$ that uses only maximizers of $f$, almost everywhere, and satisfies $\int b_d d\mc G = \beta_d$ for every $d$. 
\end{lemma}
\begin{proof}
	First, partition types by their limiting best-response sets as follows:
	\[E_J := \set{\theta \mid \argmax_{d \in \mc D} f_d(\theta) = J}, \qquad \emptyset \ne J \subseteq \mc D\]
	After taking an appropriate subsequence, all quantities $w_{J}^d := \lim_n \int_{E_J} b_d^n d \mc G$ exist, as there are only finitely many such quantities. If $d \not \in J$, it is eventually strictly suboptimal at every type in $E_J$, apart from a fixed null set, so bounded convergence yields $w_J^d = 0$. Further, we have that 
	\[\sum_d w_J^d = \mc G(E_J), \qquad \sum_J w_J^d = \beta_d\]
	On each positive measure $E_J$, we may then consider the strategy that uses constant probabilities $b_d := w_J^d/ \mc G(E_J)$. On null sets, we instead select the first maximizing action in a fixed ordering. These choices are measurable and have the required properties.
\end{proof}

We apply this Lemma twice to get the desired result. For each $(\sigma_R, q)$ we apply it to the actions accept and reject, and the receiver's payoff from accepting in the $n$th approximate game under beliefs $\lambda^n$. It yields a measurable receiver best response $a$ to $\lambda$ with $\int a(\theta, \sigma_R, q) d \mc G = \alpha_{\sigma_R, q}$. This verifies condition 1 of our sequential equilibrium concept. For each $\sigma_P$, we apply the Lemma to actions $\mc Q$, and the proposer's payoffs in the $n$th approximate game under receiver acceptance probabilities $\alpha^n_{\sigma_R, q}$. It yields a corresponding $p$ with $\int p_q(\theta, \sigma_P) d \mc G = \ell_q(\sigma_P)$ that is supported on maximizers of the proposer's true utility. This verifies condition 2. 

Finally to verify conditions 3 and 4, for each $n$ we choose a sufficiently large $k$ whose entries are within $\frac1n$ of $\ell^n$ and whose posteriors are within $\frac1n$ of $\lambda^n$. There are finitely many entries, so the choice can be made simultaneously for all signals and quotes. Since the strategy $p$ has likelihood $\ell$, these yield a valid sequence of likelihoods that satisfy conditions 3 and 4, yielding the desired result.

%% file: aimethodology.tex
\section*{AI Methodology}

We organize this section according to the template suggested recently
by the Simons Institute working group.

\begin{flushleft}
\renewcommand{\arraystretch}{1.3}
\begin{tabular}{@{}l@{\qquad}c@{}}
\textbf{Activity} & \textbf{AI used} \\
Asking the research question & [No] \\
Coming up with the approach & [Yes] \\
Proof development & [Yes] \\
Writing and exposition & [Yes] \\
Checking for bugs & [Yes] \\
Other supporting tasks & [Yes] \\
\end{tabular}
\end{flushleft}

\begin{sloppypar}
\paragraph{Details (elaborating on the points above).}
The authors developed the research questions studied in this paper and
the high-level IPV reduction approach without AI assistance. 
For writing and exposition, AI (primarily Codex) was used for 
routine copy-editing and in assisting with
the development of
some of the examples described in the paper. 
Many of the technical arguments in the paper were developed alongside
extensive AI collaboration.
The authors verified the correctness and originality of all content,
including references.
\end{sloppypar}


%% file: refs.bib
@article{M81,
        Author = {Myerson, R.},
        Journal = "Mathematics of Operations Research",
        Number = {1},
        Pages = {58--73},
        Title = {Optimal auction design},
        Volume = {6},
        Year = {1981}}

@article{NR99,
  author =  {N. Nisan and A. Ronen},
    title =      {Algorithmic Mechanism Design},
       journal = "Games and Economic Beahvior",
          year = 2001,
             volume = 35,
             number = "1--2",
                pages = {166--196}
                  }

@article{guihuang2022,
  author  = {Gui, Qingyun and Huang, Yi C.},
  title   = {On {Condorelli} inequality for log-concave demand},
  journal = {Economics Letters},
  volume  = {215},
  pages   = {110502},
  year    = {2022},
  doi     = {10.1016/j.econlet.2022.110502}
}

@article{saumardwellner2014,
  author  = {Saumard, Adrien and Wellner, Jon A.},
  title   = {Log-concavity and strong log-concavity: A review},
  journal = {Statistics Surveys},
  volume  = {8},
  pages   = {45--114},
  year    = {2014},
  doi     = {10.1214/14-SS107}
}

@article{barlowmarshallproschan1963,
  author  = {Barlow, Richard E. and Marshall, Albert W. and Proschan, Frank},
  title   = {Properties of Probability Distributions with Monotone Hazard Rate},
  journal = {The Annals of Mathematical Statistics},
  year    = {1963},
  volume  = {34},
  number  = {2},
  pages   = {375--389},
  doi     = {10.1214/aoms/1177704147}
}

@article{krepswilson1982,
  author  = {Kreps, David M. and Wilson, Robert},
  title   = {Sequential Equilibria},
  journal = {Econometrica},
  year    = {1982},
  volume  = {50},
  number  = {4},
  pages   = {863--894},
  doi     = {10.2307/1912767}
}

@article{glostenmilgrom1985,
  author  = {Glosten, Lawrence R. and Milgrom, Paul R.},
  title   = {Bid, Ask and Transaction Prices in a Specialist Market
             with Heterogeneously Informed Traders},
  journal = {Journal of Financial Economics},
  year    = {1985},
  volume  = {14},
  number  = {1},
  pages   = {71--100},
  doi     = {10.1016/0304-405X(85)90044-3}
}

@article{akerlof1970,
  author  = {Akerlof, George A.},
  title   = {The Market for ``Lemons'': Quality Uncertainty
             and the Market Mechanism},
  journal = {The Quarterly Journal of Economics},
  year    = {1970},
  volume  = {84},
  number  = {3},
  pages   = {488--500},
  doi     = {10.2307/1879431}
}

@article{mussarosen1978,
  author  = {Mussa, Michael and Rosen, Sherwin},
  title   = {Monopoly and Product Quality},
  journal = {Journal of Economic Theory},
  year    = {1978},
  volume  = {18},
  number  = {2},
  pages   = {301--317},
  doi     = {10.1016/0022-0531(78)90085-6}
}

@article{duffiegarleanupedersen2005,
  author  = {Duffie, Darrell and G{\^a}rleanu, Nicolae
             and Pedersen, Lasse Heje},
  title   = {Over-the-Counter Markets},
  journal = {Econometrica},
  year    = {2005},
  volume  = {73},
  number  = {6},
  pages   = {1815--1847},
  doi     = {10.1111/j.1468-0262.2005.00639.x}
}

@article{rostekyoon2021,
  author  = {Rostek, Marzena and Yoon, Ji Hee},
  title   = {Exchange Design and Efficiency},
  journal = {Econometrica},
  year    = {2021},
  volume  = {89},
  number  = {6},
  pages   = {2887--2928},
  doi     = {10.3982/ECTA16537}
}

@article{milgrom_weber,
	Author = {Paul R. Milgrom and Robert J. Weber},
	Journal = "Econometrica",
	Number = 5,
	Pages = {1089--1122},
	Title = {A Theory of Auctions and Competitive Bidding},
	Volume = 50,
	Year = 1982}

@article{MS83,
author = "Roger B. Myerson and Mark A. Satterthwaite",
title = "Efficient Mechanisms for Bilateral Trading",
journal = "Journal of Economic Theory",
volume = 29,
number = 2,
pages = "265--281",
year = 1983
}

@article{wilson80,
author = "Charles Wilson",
title = "The nature of equilibrium in markets with adverse selection",
journal = "The Bell Journal of Economics",
volume = 11,
number = 1,
year = 1980
}

@article{KZ26,
  author  = {Kartik, Navin and Zhong, Weijie},
    title   = {Lemonade from Lemons: Information Design and Adverse Selection},
      journal = {Theoretical Economics},
        year    = {2026},
	  volume  = {21},
	    number  = {1},
	      pages   = {281--324},
	        doi     = {10.3982/TE6451},
		  url     = {https://doi.org/10.3982/TE6451}
		  }

@article{RT16,
  author  = {Roughgarden, Tim and Talgam-Cohen, Inbal},
    title   = {Optimal and Robust Mechanism Design with Interdependent Values},
      journal = {ACM Transactions on Economics and Computation},
        year    = {2016},
	  volume  = {4},
	    number  = {3},
	      pages   = {18:1--18:34},
	        doi     = {10.1145/2910577},
		  url     = {https://doi.org/10.1145/2910577}
		  }

@article{VW11,
  author = {Voorneveld, Mark and Weibull, J{\"o}rgen W.},
    title = {A Scent of Lemon---Seller Meets Buyer with a Noisy Quality Observation},
      journal = {Games},
        year = {2011},
	  volume = {2},
	    number = {1},
	      pages = {163--186},
	        doi = {10.3390/g2010163},
		  url = {https://doi.org/10.3390/g2010163}
		  }

@article{levin01,
  author = {Levin, Jonathan},
    title = {Information and the Market for Lemons},
      journal = {The RAND Journal of Economics},
        year = {2001},
	  volume = {32},
	    number = {4},
	      pages = {657--666},
	        doi = {10.2307/2696386},
		  url = {https://doi.org/10.2307/2696386}
		  }

@unpublished{BD23,
  author = {Brooks, Benjamin and Du, Songzi},
    title = {Robust Mechanisms for the Financing of Public Goods},
      year = {2023},
        month = jun,
	  note = {Working paper, June 16, 2023},
	    url = {https://benjaminbrooks.net/downloads/brooksdu_publicgoods.pdf}
	    }

@inproceedings{D+25,
  author    = {Dobzinski, Shahar and Eden, Alon and Goldner, Kira and Shaulker, Ariel and Tsilivis, Thodoris},
    title     = {Bilateral Trade with Interdependent Values: Information vs. Approximation},
      booktitle = {Proceedings of the 26th ACM Conference on Economics and Computation},
        series    = {EC '25},
	  year      = {2025},
	    pages     = {641--665},
	      publisher = {Association for Computing Machinery},
	        doi       = {10.1145/3736252.3742607},
		  url       = {https://doi.org/10.1145/3736252.3742607}
		  }

@inproceedings{DS24,
  author    = {Dobzinski, Shahar and Shaulker, Ariel},
    title     = {Bilateral Trade with Correlated Values},
      booktitle = {Proceedings of the 56th Annual ACM Symposium on Theory of Computing},
        series    = {STOC '24},
	  year      = {2024},
	    pages     = {237--246},
	      publisher = {Association for Computing Machinery},
	        doi       = {10.1145/3618260.3649659},
		  url       = {https://doi.org/10.1145/3618260.3649659}
		  }

@inproceedings{D+22,
  author    = {Deng, Yuan and Mao, Jieming and Sivan, Balasubramanian and Wang, Kangning},
    title     = {Approximately Efficient Bilateral Trade},
      booktitle = {Proceedings of the 54th Annual ACM SIGACT Symposium on Theory of Computing},
        series    = {STOC '22},
	  year      = {2022},
	    pages     = {718--721},
	      publisher = {Association for Computing Machinery},
	        doi       = {10.1145/3519935.3520054},
		  url       = {https://doi.org/10.1145/3519935.3520054}
		  }

@inproceedings{F22,
  author    = {Fei, Yumou},
    title     = {Improved Approximation to First-Best Gains-from-Trade},
      booktitle = {Web and Internet Economics: 18th International Conference, WINE 2022},
        series    = {Lecture Notes in Computer Science},
	  volume    = {13778},
	    year      = {2022},
	      pages     = {204--218},
	        publisher = {Springer},
		  doi       = {10.1007/978-3-031-22832-2_12},
		    url       = {https://doi.org/10.1007/978-3-031-22832-2_12}
		    }

@misc{BWZ26,
  author        = {Bei, Xiaohui and Wu, Wenhao and Zhou, Shengwei},
    title         = {The Power of Simple Mechanisms: A Tight {$e$}-Approximation for Gains from Trade in Matching Markets},
      year          = {2026},
        month         = sep,
	  eprint        = {2609.37997},
	    archivePrefix = {arXiv},
	      primaryClass  = {cs.GT},
	        note          = {Preprint, arXiv:2609.37997},
		  url           = {https://arxiv.org/abs/2609.37997}
		  }

@misc{J26,
  author        = {Jo, Sunghyeon},
    title         = {Tighter bounds for the random-offerer mechanism in bilateral trade},
      year          = {2026},
        month         = jul,
	  eprint        = {2607.13959},
	    archivePrefix = {arXiv},
	      primaryClass  = {cs.GT},
	        note          = {Preprint, arXiv:2607.13959},
		  url           = {https://arxiv.org/abs/2607.13959}
		  }

@article{E+19,
  author  = {Eden, Alon and Feldman, Michal and Fiat, Amos and Goldner, Kira and Karlin, Anna R.},
    title   = {Combinatorial Auctions with Interdependent Valuations: {SOS} to the Rescue},
      journal = {Mathematics of Operations Research},
        year    = {2024},
	  volume  = {49},
	    number  = {2},
	      pages   = {653--674},
	        doi     = {10.1287/moor.2023.1371},
		  url     = {https://doi.org/10.1287/moor.2023.1371}
		  }

@inproceedings{E+22,
  author    = {Eden, Alon and Goldner, Kira and Zheng, Shuran},
    title     = {Private Interdependent Valuations},
      booktitle = {Proceedings of the 2022 Annual ACM-SIAM Symposium on Discrete Algorithms (SODA)},
        year      = {2022},
	  pages     = {2920--2939},
	    publisher = {Society for Industrial and Applied Mathematics},
	      doi       = {10.1137/1.9781611977073.113},
	        url       = {https://doi.org/10.1137/1.9781611977073.113}
		}

@inproceedings{E+23,
  author    = {Eden, Alon and Feldman, Michal and Goldner, Kira and Mauras, Simon and Mohan, Divyarthi},
    title     = {Constant Approximation for Private Interdependent Valuations},
      booktitle = {2023 IEEE 64th Annual Symposium on Foundations of Computer Science (FOCS)},
        year      = {2023},
	  pages     = {148--163},
	    publisher = {IEEE Computer Society},
	      doi       = {10.1109/FOCS57990.2023.00018},
	        url       = {https://doi.org/10.1109/FOCS57990.2023.00018}
		}

@inproceedings{B+17,
  author    = {Brustle, Johannes and Cai, Yang and Wu, Fa and Zhao, Mingfei},
    title     = {Approximating Gains from Trade in Two-sided Markets via Simple Mechanisms},
      booktitle = {Proceedings of the 2017 ACM Conference on Economics and Computation},
        series    = {EC '17},
	  year      = {2017},
	    pages     = {589--590},
	      publisher = {Association for Computing Machinery},
	        doi       = {10.1145/3033274.3085148},
		  url       = {https://doi.org/10.1145/3033274.3085148}
		  }

@inproceedings{C+21,
  author    = {Cai, Yang and Goldner, Kira and Ma, Steven and Zhao, Mingfei},
    title     = {On Multi-Dimensional Gains from Trade Maximization},
      booktitle = {Proceedings of the 2021 ACM-SIAM Symposium on Discrete Algorithms (SODA)},
        year      = {2021},
	  pages     = {1079--1098},
	    publisher = {Society for Industrial and Applied Mathematics},
	      doi       = {10.1137/1.9781611976465.67},
	        url       = {https://doi.org/10.1137/1.9781611976465.67}
		}

@inproceedings{B+26,
  author    = {Babaioff, Moshe and Rubinstein, Aviad and Tan, Xizhi and Wang, Kangning},
    title     = {Approximating Gains-from-Trade in Matching Markets},
      booktitle = {Proceedings of the 58th Annual ACM Symposium on Theory of Computing},
        series    = {STOC '26},
	  year      = {2026},
	    pages     = {710--721},
	      publisher = {Association for Computing Machinery},
	        doi       = {10.1145/3798129.3800786},
		  url       = {https://doi.org/10.1145/3798129.3800786}
		  }

@article{E+21,
  author  = {Eden, Alon and Feldman, Michal and Talgam-Cohen, Inbal and Zviran, Ori},
    title   = {{PoA} of Simple Auctions with Interdependent Values},
      journal = {Proceedings of the AAAI Conference on Artificial Intelligence},
        year    = {2021},
	  volume  = {35},
	    number  = {6},
	      pages   = {5321--5329},
	        doi     = {10.1609/aaai.v35i6.16671},
		  url     = {https://ojs.aaai.org/index.php/AAAI/article/view/16671}
		  }

@inproceedings{PST12,
  author    = {Paes Leme, Renato and Syrgkanis, Vasilis and Tardos, {\'E}va},
    title     = {Sequential Auctions and Externalities},
      booktitle = {Proceedings of the Twenty-Third Annual ACM-SIAM Symposium on Discrete Algorithms},
        year      = {2012},
	  pages     = {869--886},
	    publisher = {Society for Industrial and Applied Mathematics},
	      doi       = {10.1137/1.9781611973099.70},
	        url       = {https://doi.org/10.1137/1.9781611973099.70}
		}

@inproceedings{ST12,
  author    = {Syrgkanis, Vasilis and Tardos, {\'E}va},
    title     = {{Bayesian} Sequential Auctions},
      booktitle = {Proceedings of the 13th ACM Conference on Electronic Commerce},
        year      = {2012},
	  pages     = {929--944},
	    publisher = {Association for Computing Machinery},
	      doi       = {10.1145/2229012.2229082},
	        url       = {https://arxiv.org/abs/1206.4771}
		}

@inproceedings{B+23,
  author    = {Babaioff, Moshe and Immorlica, Nicole and Li, Yingkai and Lucier, Brendan},
    title     = {Making Auctions Robust to Aftermarkets},
      booktitle = {14th Innovations in Theoretical Computer Science Conference (ITCS 2023)},
        year      = {2023},
	  series    = {Leibniz International Proceedings in Informatics},
	    volume    = {251},
	      pages     = {9:1--9:23},
	        publisher = {Schloss Dagstuhl -- Leibniz-Zentrum f{\"u}r Informatik},
		  doi       = {10.4230/LIPIcs.ITCS.2023.9},
		    url       = {https://drops.dagstuhl.de/entities/document/10.4230/LIPIcs.ITCS.2023.9}
		    }

@inproceedings{B+26b,
  author    = {Banerjee, Siddhartha and Munagala, Kamesh and Shen, Yiheng and Wang, Kangning},
    title     = {The Price of Competitive Information Disclosure},
      booktitle = {Proceedings of the 58th Annual ACM Symposium on Theory of Computing},
        year      = {2026},
	  pages     = {1302--1313},
	    publisher = {Association for Computing Machinery},
	      doi       = {10.1145/3798129.3800840},
	        url       = {https://doi.org/10.1145/3798129.3800840}
		}

@inproceedings{FLS13,
  author    = {Feldman, Michal and Lucier, Brendan and Syrgkanis, Vasilis},
    title     = {Limits of Efficiency in Sequential Auctions},
      booktitle = {Web and Internet Economics: 9th International Conference, WINE 2013},
        year      = {2013},
	  series    = {Lecture Notes in Computer Science},
	    volume    = {8289},
	      pages     = {160--173},
	        publisher = {Springer},
		  doi       = {10.1007/978-3-642-45046-4_14},
		    url       = {https://doi.org/10.1007/978-3-642-45046-4_14}
		    }

@inproceedings{DPT16,
  author    = {Daskalakis, Constantinos and Papadimitriou, Christos H. and Tzamos, Christos},
    title     = {Does Information Revelation Improve Revenue?},
      booktitle = {Proceedings of the 2016 ACM Conference on Economics and Computation},
        year      = {2016},
	  pages     = {233--250},
	    publisher = {Association for Computing Machinery},
	      doi       = {10.1145/2940716.2940789},
	        url       = {https://doi.org/10.1145/2940716.2940789}
		}

@inproceedings{BD14,
  author    = {Blumrosen, Liad and Dobzinski, Shahar},
  title     = {Reallocation Mechanisms},
  booktitle = {Proceedings of the Fifteenth {ACM} Conference on Economics
               and Computation},
  year      = {2014},
  pages     = {617},
  publisher = {Association for Computing Machinery},
  doi       = {10.1145/2600057.2602843},
  url       = {https://doi.org/10.1145/2600057.2602843}
}

@article{BD21,
  author  = {Blumrosen, Liad and Dobzinski, Shahar},
  title   = {(Almost) Efficient Mechanisms for Bilateral Trading},
  journal = {Games and Economic Behavior},
  year    = {2021},
  volume  = {130},
  pages   = {369--383},
  doi     = {10.1016/j.geb.2021.08.011},
  url     = {https://doi.org/10.1016/j.geb.2021.08.011}
}

@inproceedings{KPV22,
  author    = {Kang, Zi Yang and Pernice, Francisco and Vondr{\'a}k, Jan},
  title     = {Fixed-Price Approximations in Bilateral Trade},
  booktitle = {Proceedings of the 2022 Annual {ACM-SIAM} Symposium
               on Discrete Algorithms ({SODA})},
  year      = {2022},
  pages     = {2964--2985},
  publisher = {Society for Industrial and Applied Mathematics},
  doi       = {10.1137/1.9781611977073.115},
  url       = {https://doi.org/10.1137/1.9781611977073.115}
}

@inproceedings{CW23,
  author    = {Cai, Yang and Wu, Jinzhao},
  title     = {On the Optimal Fixed-Price Mechanism in Bilateral Trade},
  booktitle = {Proceedings of the 55th Annual {ACM} Symposium
               on Theory of Computing},
  year      = {2023},
  pages     = {737--750},
  publisher = {Association for Computing Machinery},
  doi       = {10.1145/3564246.3585171},
  url       = {https://doi.org/10.1145/3564246.3585171}
}

@inproceedings{LRW23,
  author    = {Liu, Zhengyang and Ren, Zeyu and Wang, Zihe},
  title     = {Improved Approximation Ratios of Fixed-Price Mechanisms
               in Bilateral Trades},
  booktitle = {Proceedings of the 55th Annual {ACM} Symposium
               on Theory of Computing},
  year      = {2023},
  pages     = {751--760},
  publisher = {Association for Computing Machinery},
  doi       = {10.1145/3564246.3585160},
  url       = {https://doi.org/10.1145/3564246.3585160}
}

@article{GdK26,
  author  = {Giambartolomei, Giordano and de Keijzer, Bart},
  title   = {On the Approximation Ratio of Optimal Fixed-Price Mechanisms
             for Single and Multi-Unit Bilateral Trade},
  journal = {Proceedings of the {AAAI} Conference on Artificial Intelligence},
  year    = {2026},
  volume  = {40},
  number  = {20},
  pages   = {16946--16953},
  doi     = {10.1609/aaai.v40i20.38741},
  url     = {https://doi.org/10.1609/aaai.v40i20.38741}
}

@inproceedings{C+17,
  author    = {Colini-Baldeschi, Riccardo and Goldberg, Paul
               and de Keijzer, Bart and Leonardi, Stefano
               and Turchetta, Stefano},
  title     = {Fixed Price Approximability of the Optimal Gain from Trade},
  booktitle = {Web and Internet Economics: 13th International Conference,
               {WINE} 2017, Proceedings},
  editor    = {Devanur, Nikhil R. and Lu, Pinyan},
  series    = {Lecture Notes in Computer Science},
  volume    = {10660},
  year      = {2017},
  pages     = {146--160},
  publisher = {Springer},
  doi       = {10.1007/978-3-319-71924-5_11},
  url       = {https://doi.org/10.1007/978-3-319-71924-5_11}
}

@inproceedings{BM16,
  author    = {Blumrosen, Liad and Mizrahi, Yehonatan},
  title     = {Approximating Gains-from-Trade in Bilateral Trading},
  booktitle = {Web and Internet Economics: 12th International Conference,
               {WINE} 2016, Proceedings},
  editor    = {Cai, Yang and Vetta, Adrian},
  series    = {Lecture Notes in Computer Science},
  volume    = {10123},
  year      = {2016},
  pages     = {400--413},
  publisher = {Springer},
  doi       = {10.1007/978-3-662-54110-4_28},
  url       = {https://doi.org/10.1007/978-3-662-54110-4_28}
}

@article{M08,
  author  = {McAfee, R. Preston},
  title   = {The Gains from Trade under Fixed Price Mechanisms},
  journal = {Applied Economics Research Bulletin},
  volume  = {1},
  number  = {1},
  pages   = {1--10},
  year    = {2008},
  url     = {https://prestonmcafee.net/PDF/GFTConstantPrice.pdf}
}

@unpublished{KV19,
  author = {Kang, Zi Yang and Vondr{\'a}k, Jan},
  title  = {Fixed-Price Approximations to Optimal Efficiency
            in Bilateral Trade},
  year   = {2019},
  note   = {Working paper},
  doi    = {10.2139/ssrn.3460336},
  url    = {https://ssrn.com/abstract=3460336}
}

@inproceedings{B+18,
  author    = {Babaioff, Moshe and Cai, Yang and
               Gonczarowski, Yannai A. and Zhao, Mingfei},
  title     = {The Best of Both Worlds: Asymptotically Efficient
               Mechanisms with a Guarantee on the Expected
               Gains-From-Trade},
  booktitle = {Proceedings of the 2018 {ACM} Conference
               on Economics and Computation},
  series    = {EC '18},
  year      = {2018},
  pages     = {373},
  publisher = {Association for Computing Machinery},
  doi       = {10.1145/3219166.3219203},
  url       = {https://doi.org/10.1145/3219166.3219203}
}

@inproceedings{CFK14,
  author    = {Chawla, Shuchi and Fu, Hu and Karlin, Anna R.},
  title     = {Approximate Revenue Maximization in Interdependent
               Value Settings},
  booktitle = {Proceedings of the Fifteenth {ACM} Conference
               on Economics and Computation},
  series    = {EC '14},
  year      = {2014},
  pages     = {277--294},
  publisher = {Association for Computing Machinery},
  doi       = {10.1145/2600057.2602858},
  url       = {https://doi.org/10.1145/2600057.2602858}
}

@article{L17,
  author  = {Li, Yunan},
  title   = {Approximation in Mechanism Design with
             Interdependent Values},
  journal = {Games and Economic Behavior},
  year    = {2017},
  volume  = {103},
  pages   = {225--253},
  doi     = {10.1016/j.geb.2016.01.006},
  url     = {https://doi.org/10.1016/j.geb.2016.01.006}
}

@inproceedings{E+18,
  author    = {Eden, Alon and Feldman, Michal and Fiat, Amos
               and Goldner, Kira},
  title     = {Interdependent Values without Single-Crossing},
  booktitle = {Proceedings of the 2018 {ACM} Conference
               on Economics and Computation},
  series    = {EC '18},
  year      = {2018},
  pages     = {369},
  publisher = {Association for Computing Machinery},
  doi       = {10.1145/3219166.3219173},
  url       = {https://doi.org/10.1145/3219166.3219173}
}

@misc{B+21,
  author        = {Babaioff, Moshe and Dobzinski, Shahar and Kupfer, Ron},
  title         = {A Note on the Gains from Trade of the Random-Offerer Mechanism},
  year          = {2021},
  eprint        = {2111.07790},
  archivePrefix = {arXiv},
  primaryClass  = {cs.GT},
  url           = {https://arxiv.org/abs/2111.07790}
}

@book{shaked2007stochastic,
  author    = {Shaked, Moshe and Shanthikumar, J. George},
  title     = {Stochastic Orders},
  series    = {Springer Series in Statistics},
  publisher = {Springer},
  address   = {New York},
  year      = {2007},
  doi       = {10.1007/978-0-387-34675-5}
}

@article{roughgarden2015intrinsic,
  author  = {Roughgarden, Tim},
  title   = {Intrinsic Robustness of the Price of Anarchy},
  journal = {Journal of the ACM},
  volume  = {62},
  number  = {5},
  pages   = {32:1--32:42},
  year    = {2015},
  doi     = {10.1145/2806883}
}

@article{KG11,
  author  = {Kamenica, Emir and Gentzkow, Matthew},
  title   = {{Bayesian} Persuasion},
  journal = {American Economic Review},
  year    = {2011},
  volume  = {101},
  number  = {6},
  pages   = {2590--2615},
  doi     = {10.1257/aer.101.6.2590},
  url     = {https://doi.org/10.1257/aer.101.6.2590}
}

@article{AAB20,
  author  = {Abraham, Ittai and Athey, Susan and Babaioff, Moshe
             and Grubb, Michael D.},
  title   = {Peaches, Lemons, and Cookies: Designing Auction Markets
             with Dispersed Information},
  journal = {Games and Economic Behavior},
  year    = {2020},
  volume  = {124},
  pages   = {454--477},
  doi     = {10.1016/j.geb.2020.09.004},
  url     = {https://doi.org/10.1016/j.geb.2020.09.004}
}
